\documentclass[aps,showpacs]{revtex4}
\newcommand{\cd}{\makebox[0.08cm]{$\cdot$}}

\usepackage{epsf}     
 
\begin{document}

\title{Relativistic bound states in Yukawa model}

\author{M. Mangin-Brinet and J. Carbonell}
\affiliation{Institut des Sciences Nucl\'{e}aires, 
        53, Av. des Martyrs, 38026 Grenoble, France} 
\author{V.A. Karmanov}
\affiliation{Lebedev Physical Institute, Leninsky Pr. 53, 
119991 Moscow, Russia}%

\date{\today}

\begin{abstract}{The  bound state solutions of two fermions interacting 
by a scalar exchange are obtained in the framework of the explicitly 
covariant light-front dynamics.  The stability with respect to cutoff 
of the J$^{\pi}$=$0^+$  and J$^{\pi}$=$1^+$ states is studied.  The 
solutions for J$^{\pi}$=$0^+$ are found to be stable for coupling 
constants $\alpha={g^2\over4\pi}$ below the critical value 
$\alpha_c\approx 3.72$ and unstable above it.  The asymptotic behavior 
of the wave functions is found to follow a ${1\over k^{2+\beta}}$ law.  
The coefficient $\beta$ and the critical coupling constant $\alpha_c$ 
are calculated from an eigenvalue equation.  The binding energies for 
the J$^{\pi}$=$1^+$ solutions diverge logarithmically with the cutoff 
for any value of the coupling constant. 
For a wide range of cutoff, the states with different 
angular momentum projections are weakly split.} \end{abstract}


\maketitle

\section{Introduction}

One of the most difficult problems in field theory is the calculation 
of bound states due to the fact that they necessarily involve an 
infinite number of diagrams.  A promising approach to deal with this 
problem is Light-Front Dynamics (LFD).  In its standard version 
\cite{BPP_PR_98}, the state vector is defined on the surface $t+z=0$.  
The bound states in the Yukawa model (two fermions interacting by 
scalar exchange) were studied in references \cite{glazek1,glazek2} 
using Tamm-Dancoff method (see, e.g., \cite{perry,Bakker}).  It was 
found in particular that, because the dominating kernel at large 
momenta tends to a constant, the binding energy of the $J=0^+$ state is 
cutoff dependent, what requires the renormalization of the Hamiltonian.

The two-fermion wave functions were also considered in the explicitly 
covariant version of LFD (CLFD) \cite{cdkm}.  In this formalism, 
proposed in \cite{karm76}, the state vector is defined on the plane 
given by the invariant equation $\omega\cd x=0$ with $\omega^2=0$.  
This approach keeps all along explicitly the dependence of the 
amplitudes on the light-front normal $\vec{\omega}$ and presents some 
advantages, in particular when calculating form factors \cite{SK} and 
 when constructing non zero angular momentum states \cite{heidelberg}.  
A first attempt to deal with CLFD wave functions was done in 
\cite{ckj1,ckj0} where deuteron and $pn$ scattering $J^{\pi}=0^+$ state 
were calculated perturbatively and successfully applied to deuteron 
e.m.  form factors \cite{ck-epj} measured at TJNAF \cite{t20}.

The CLFD equations have now been solved exactly 
for a two-fermion  system in the ladder approximation with different 
boson exchange couplings \cite{These_MMB,fermions}.  The first results 
for the Yukawa model have been reported in \cite{mck_prd}.  We 
investigated with special interest the stability of the bound state 
solutions relative to the cutoff, disregarding the self energy 
contribution and renormalization.  We have found a critical phenomenon 
for the cutoff dependence of the binding energy. The $J=0^+$ solutions 
were found to be stable -- i.e. with finite limit when the cutoff tends 
to infinity -- for coupling constant $\alpha$  below a critical value 
$\alpha_c$. On the contrary, for values exceeding  $\alpha_c$, the 
system collapses. This fact manifests itself either as an infinite 
number of bound states with 
unbounded energies going to $-\infty$ for a finite value of the 
coupling constant or as a zero value of the coupling constant for a 
fixed value of the binding energy.

The present paper is a detailed version of our work \cite{mck_prd}, 
includes new findings concerning the calculation of the critical 
coupling constant and the asymptotical behavior of the wave functions
and the full treatment of the $J=1^+$ state. Our results are compared 
to those obtained in \cite{glazek1}.

In section \ref{covwf} we remind the general properties of the 
two-fermion equations and wave functions in CLFD, already presented in 
\cite{cdkm}.  In sections \ref{J0} and \ref{J1} the system of equations 
for the $J^{\pi}=0^+$ and $J^{\pi}=1^+$ wave function components are 
derived.  We show in section \ref{relgl} that, after a linear 
transformation of these components and a change of variables, the CLFD 
equations are identical to the ones considered in \cite{glazek1}.  In 
section \ref{asympt} we analyze analytically the asymptotical 
properties of the kernels and wave functions and their relation with the 
existence of a critical coupling constant.  Numerical results are 
presented in section \ref{num} and section \ref{concl} contains the 
concluding remarks.

\section{Covariant wave function and equation}\label{covwf}

We briefly describe here the main properties of the CLFD wave functions 
and equations. A detailed derivation can be found in 
\cite{cdkm,These_MMB,fermions}. 

In the covariant version of LFD the wave functions are the Fock 
components of the state vector defined on the light-front plane 
$\omega\cd x=0$.  The standard LFD approach is recovered as a 
particular case with $\omega=(1,0,0,-1)$.

\begin{figure}[h]
\centerline{\epsfbox{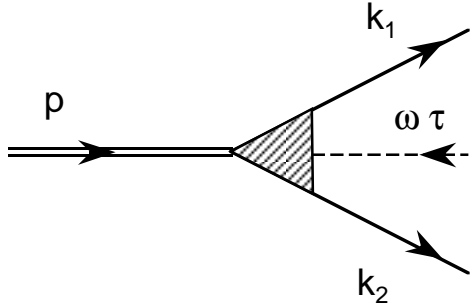}}
\caption{Graphical representation of the two-body wave function.}\label{fwf}
\end{figure}
The wave function ${\mit \Phi}_{\sigma_2\sigma_1}$ of a two-fermion bound state 
is shown graphically in figure \ref{fwf}.
It depends on four four-momenta
\begin{equation}\label{wf1}
{\mit \Phi}_{\sigma_2\sigma_1}={\mit \Phi}_{\sigma_2\sigma_1}(k_1,k_2,p,\omega\tau)
\end{equation}
which are all on the corresponding mass shells 
($k_1^2=k_2^2=m^2$, $p^2=M^2$, $(\omega\tau)^2=0$) and satisfy the conservation law: 
\begin{equation}\label{kcl} 
k_1+k_2=p+\omega\tau.
\end{equation}

In the standard approach the $+$ and $\perp$ components are conserved, 
whereas the minus-component is not.  These properties are reproduced by 
equation (\ref{kcl}), since in this case the only nonzero component is 
$\omega_-=2$.  Equation (\ref{kcl}) is thus a covariant generalization 
of the usual conservation law.

The general form of the  wave function (\ref{wf1}) depends on the 
particular quantum number of the state and is obtained by constructing 
all possible spin structures. 

It is convenient to introduce variables ($\vec{k},\hat{n}$), 
constructed from the initial four-momenta as follows:
\begin{eqnarray}                                                   
\vec{k}&=& L^{-1}({\cal P})\vec{k}_1 = \vec{k}_1 -                                
\frac{\vec{\cal P}}{\sqrt{{\cal P}^2}}[k_{10}                                   
- \frac{\vec{k}_1\cd\vec{{\cal P}}}{\sqrt{{\cal P}^2}+{\cal P}_0}]          
\nonumber \\\label{sc4}  
\hat{n}&=& {L^{-1}({\cal P})\vec{\omega}\over|L^{-1}({\cal P})\vec{\omega}|}
\end{eqnarray}                                                                  
where                                                                           
${\cal P} = p + \omega\tau$,                                     
and $L^{-1}({\cal P})$ is the Lorentz boost into the reference system where
$\vec{\cal P}=0$. 
In these variables, the wave function (\ref{wf1}) is represented as:
\begin{equation}\label{wf3}
{\mit \Phi}_{\sigma_2\sigma_1}={\mit \Phi}_{\sigma_2\sigma_1}(\vec{k},\hat{n}).
\end{equation}
Under Lorentz transformations of the four-momenta 
$k_1,k_2,p,\omega\tau$ the variables $\vec{k},\vec{n}$ are only rotated 
\cite{cdkm}, so that parametrization (\ref{wf3}), though being 
three-dimensional, is also explicitly covariant.  In practice, instead 
of dealing with transformations (\ref{sc4}), it is enough to write the 
wave functions and dynamical equations in the center of mass reference 
system, i.e. the one for which $\vec{\cal P}=\vec{k}_1+\vec{k}_2=0$ and 
in which $\vec{k}_1\equiv\vec{k}$, $\vec{k}_2\equiv-\vec{k}$, 
$\vec{\omega}\equiv\hat{n}|\vec{\omega}|$.

The equation for the wave function in terms of variables (\ref{sc4}) reads:
\begin{equation}\label{eq5d} 
\left[M^2-4(\vec{k}\,^2+m^2)\right]{\mit\Phi}_{\sigma_2\sigma_1}(\vec{k},\vec{n})=
\frac{m^2}{2\pi^3} \int \sum_{\sigma'_1\sigma'_2}
K_{\sigma_2\sigma_1}^{\sigma'_2\sigma'_1}(\vec{k},\vec{k}\,',\hat{n},M^2)
{\mit \Phi}_{\sigma'_2\sigma'_1}(\vec{k}\,',\hat{n}) \frac{d^3k'}{\varepsilon_{k'}}
\end{equation}
where $K(\vec{k},\vec{k}\,',\hat{n},M^2)$ is the interaction kernel.  
It depends on the scalar products of the vectors 
$\vec{k}\,',\vec{k},\vec{n}$ and also on the scalar products 
$\vec{k}\cd\vec{\sigma}$, $\vec{k}\,'\cd\vec{\sigma}$ and $\hat{n}\cd 
\vec{\sigma}$.  For the Yukawa model it will be precised in next 
section. 

In CLFD, the construction of states with definite angular momentum has 
some peculiarities which are explained in \cite{cdkm,fermions}. These 
peculiarities are related to the fact that the angular momentum 
operator $\vec{J}$ is not kinematical, but contains the interaction.  
However, we can overcome this difficulty by taking into account the 
so-called angular condition, derived from the transformation properties 
of the wave function under rotations of the light-front plane. Assuming 
that the state vector satisfies this condition, the problem results in 
finding the eigenfunctions of a purely kinematical operator $\vec{M}$ 

\begin{equation}\label{ac3}                                                     
\vec{M}=-i\; \vec{k}\times {\partial\over \partial\vec{k}}\,
-i\;\vec{n}\times                   
{\partial\over \partial\vec{n}}+\vec{s}_1+\vec{s}_2\                                
\end{equation}   
where $\vec{s}_{i}$ are the fermion spin operators. The operators  ($\vec{M}^2$, 
$M_z$) have the same eigenvalues $J(J+1)$ and  $\lambda$ than the full angular
momentum operators ($\vec{J}\,^2$, $J_z$). 

As already mentioned, in any Lorentz transformation of the state vector, 
$\vec{k}$ and $\hat{n}$ undergo only rotations, with the same rotation 
operator than the one acting on spinor indices $\sigma_1,\sigma_2$.  In 
this respect, the eigenstates of $\vec{M}^2, M_z$ are constructed as in 
the non relativistic quantum mechanics.  The only difference is that we 
have at our disposal two three-dimensional vectors ($\vec{k},\hat{n}$) 
which enter in this construction on equal ground, instead of the only 
relative momentum $\vec{k}$ in the non relativistic case. 

Since the interaction kernel in (\ref{eq5d}) depends on the scalar 
products of all the three-vectors, including  spin operator, $\vec{M}$ 
commutes with it. The solutions of the equation (\ref{eq5d}) are 
eigenfunctions of the operators $\vec{M}^2,M_z$ with eigenvalues 
$J(J+1)$ and $\lambda$.

Although the operator $\vec{M}$ contains the derivatives both over 
$\partial/\partial\vec{k}$ and $\partial/\partial\hat{n}$, 
its projection on $\hat{n}$-axis does not involve
any derivative with respect to variable $\hat{n}$.
Furthermore this latter enters in  equation (\ref{eq5d}) as a vector parameter
only and  not as a dynamical variable.
Therefore, there exists another operator which commutes with the kernel, namely:
\begin{equation}\label{ac6}                                                     
A^2=(\hat{n}\cd\vec{M})^2.
\end{equation}
Since $A^2$ is a scalar, it commutes also with $\vec{M}$.
Therefore, in addition to $J,J_z$, the solutions are labeled by $a$:
\begin{equation}\label{eqf1}
A^2\vec{\psi}^{a}(\vec{k},\hat{n})=a^2\vec{\psi}^{a}(\vec{k},\hat{n}).
\end{equation}
The operator $A^2$ has $J+1$ eigenvalues, $a^2=0,1,\ldots,J^2$, and 
$2J+1$ eigenfunctions which are split in two families of $J$ and $J+1$ 
states with opposite parities.  For $J=0$, there is only one value 
$a=0$. For $J^\pi=1^+$, there are two values $a=0$ and $a=1$.  The wave 
function with definite $J$ is determined by $N_J={1\over2}(2J+1)\times 
2\times 2=2(2J+1)$ spin components (the ${1\over2}$ factor comes from 
parity conservation).  The system of equations for these $2(2J+1)$ spin 
components is split in different subsystems each of them with definite 
value of $a$.  For example, the wave function for $J=1$ is determined 
by $N_J=6$ components \cite{ckj1} and the equation system is split in 
two subsystems corresponding to $a=0$ and $a=1$ and containing 
respectively 2 and 4 equations (see section \ref{J1}).

\begin{figure}[h]
\centerline{\epsfbox{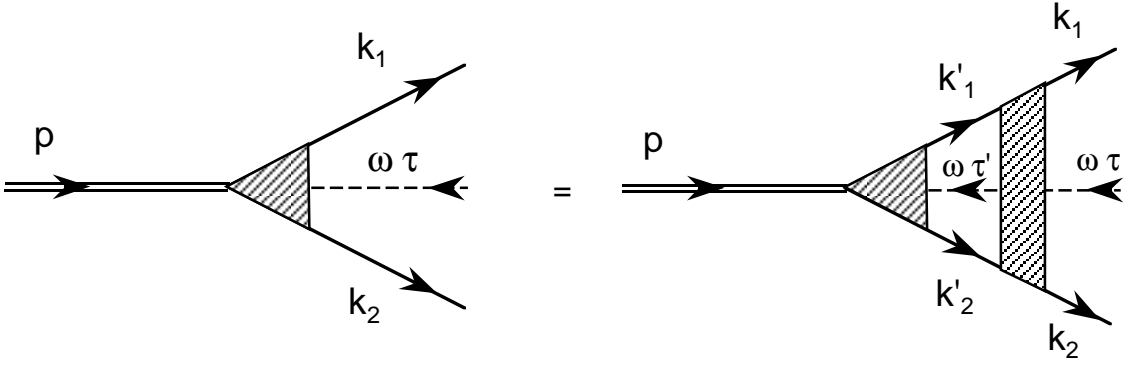}}
\caption{Equation for the two-body wave function.\label{feq}}
\end{figure}

The calculation technique of the CLFD is given by special graph rules 
which are a covariant generalization of the old fashioned perturbation 
theory. It was developed by Kadyshevsky \cite{kadysh} and adapted to 
CLFD in \cite{karm76,cdkm}.  The equation  for the wave function is 
shown graphically in figure \ref{feq} and the corresponding analytical 
form (\ref{eq5d}) is obtained by applying the rules of the graph 
techniques to the diagram displayed in this figure. 

All along this paper we will consider the one boson exchange 
kernel with scalar coupling only. The
interaction Lagrangian is  ${\cal L}^{int}=g\;\bar{\psi}\psi\phi^{(s)}$.
\begin{figure}[h]
\centerline{\epsfbox{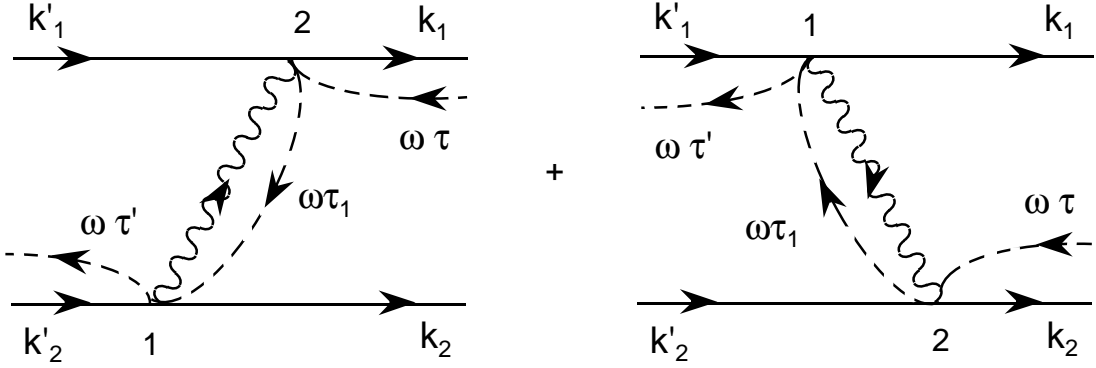}}
\caption{One boson exchange kernel.\label{fkern}}
\end{figure}
The corresponding amplitude -- represented graphically
 in figure \ref{fkern} has the form:
\begin{eqnarray}\label{4.2}                                                     
K^{\sigma'_2\sigma'_1}_{\sigma_2\sigma_1}                                       
&=&-\frac{g^2}{4m^2}
  \left[\bar{u}^{\sigma_2}(k_2) u^{\sigma'_2}(k'_2)\right]
\;\left[\bar{u}^{\sigma_1}(k_1) u^{\sigma'_1}(k'_1)\right]\\
&&\times
\left\{\frac{\theta\left(\omega\cd (k'_1 - k_1)\right)}{\mu^2- 
(k'_1-k_1)^2 + 2\tau'\omega\cd (k'_1-k_1)} \right. 
+\left. \frac{\theta\left(\omega\cd (k_1-k'_1)\right)} {\mu^2 - 
(k_1-k'_1)^2 + 2\tau\omega\cd (k_1-k'_1)}\right\}.                                                                       
\nonumber
\end{eqnarray}  
Like the wave function,
the kernel is off energy shell for $\tau,\tau'\neq 0$. 
The two terms can be simplified into:
\begin{equation}\label{eq6}                                                     
K(k_1,k_2,\omega\tau; k'_1,k'_2,\omega\tau') =-\frac{g^2}{4m^2(Q^2+\mu^2)}
\left[\bar{u}_{\sigma_2}(k_2) u_{\sigma'_2}(k'_2)\right]\,
\left[\bar{u}_{\sigma_1}(k_1) u_{\sigma'_1}(k'_1)\right],                                                                      
\end{equation}
$Q^2$ is given in variables $\vec{k},\hat{n}$ by
\begin{eqnarray}\label{k4} 
Q^2 &=& k^2+k'^2
-2kk'\left(1+\frac{(\varepsilon_k -\varepsilon_{k'})^2}
{2\varepsilon_k\varepsilon_{k'}}\right)\cos\theta\cos\theta'
-2k k'\sin\theta \sin\theta'\cos\varphi'
\nonumber\\
&&+\left(\varepsilon_{k}^2
+\varepsilon_{k'}^2-\frac{1}{2}M^2\right)                           
\left|\frac{k\cos\theta}{\varepsilon_{k}}                                
-\frac{k'\cos\theta'}{\varepsilon_{k'}}\right|                                 
\end{eqnarray}
where $k\cos\theta=\hat{n}\cd\vec{k}$, $k'\cos\theta'=\hat{n}\cd\vec{k}'$ 
and $\varphi'$ is the azimuthal angle between $\vec{k}$ and 
$\vec{k}\,'$ in the plane orthogonal to $\hat{n}$.

\section{The $J=0^+$ state}\label{J0}

The wave function of a two fermion system in the state $J=0^+$ has the form 
\cite{ckj0}:
\begin{equation}\label{eq1} 
{\mit \Phi}_{\sigma_2\sigma_1}(k_1,k_2,p,\omega\tau)=
\sqrt{m}\overline{u}_{\sigma_2}(k_2)
\phi U_c \overline{u}_{\sigma_1}(k_1),
\end{equation}
with
\begin{equation}\label{eq1_1} 
\phi=f_1S_1 +  f_2 S_2,
\end{equation}
$f_{i}$ are scalar functions depending on $(k,\theta)$ and $S_{i}$ are 
spin structures given by 
\begin{equation}\label{eq1_2} 
S_1=\frac{1}{2\sqrt{2}\varepsilon_k}\gamma_5,\quad
S_2=\frac{\varepsilon_k}{2\sqrt{2} mk\sin\theta}
\left(\frac{2m\hat{\omega}}{\omega\cd p}- 
\frac{m^2}{\varepsilon_k^2}\right)\gamma_5 
\end{equation}
in which $\hat{\omega}=\omega_\mu\gamma^\mu$ and $\varepsilon_{k}=\sqrt{k^2+m^2}$.
$U_c=\gamma^2 \gamma^0$ is the charge conjugation matrix,
$u$ the usual spinor normalized as 
$\bar{u}_{\sigma}u_{\sigma'}=2m\delta_{\sigma\sigma'}$:
\begin{equation}\label{eq74} 
u_{\sigma}(k)= \sqrt{\varepsilon_{k}+m} \left(\begin{array}{c} 1\\
\frac{\displaystyle\vec{\sigma}\cd\vec{k}}{\displaystyle(\varepsilon_{k}+m)}
\end{array}\right) w_{\sigma}, 
\end{equation}
and $w_{\sigma}$ the two-component spinor normalized to
$w^{\dagger}_{\sigma}w_{\sigma'}=\delta_{\sigma\sigma'}$.  

Because of the spin structure $\hat{\omega}$, the relativistic wave 
function (\ref{eq1}) is determined by two scalar functions, in the 
agreement with the counting rule mentioned above. In section 
 \ref{relgl} we will show that this state corresponds in the standard 
approach to the $(1+,2-)$ state discussed in \cite{glazek1} and 
described by the two components $\Phi^{1+},\Phi^{2-}$. 

In the reference system where $\vec{k}_1+\vec{k}_2=0$, the 
wave function (\ref{eq1}) can be represented as follows: 
\[{\mit \Phi}_{\sigma_2\sigma_1}=\sqrt{m}
w_{\sigma_2}^{\dagger}\psi(\vec{k},\vec{n})w_{\sigma_1}^{\dagger}\]
with
\begin{equation}\label{eq0} 
\psi(\vec{k},\hat{n})=\frac{1}{\sqrt{2}}
\left(f_1+\frac{i\vec{\sigma}\cd [\vec{k}\times\hat{n}]}{k\sin\theta}f_2\right)\sigma_y
\end{equation}

The normalization condition has the form:
\begin{eqnarray}\label{na5} 
&&\sum_{\sigma_2\sigma_1}\left|{\mit \Phi}_{\sigma_2\sigma_1}\right|^2=
{m\over (2\pi)^3}\int Tr\{\bar{\phi}(\hat{k}_2+m) 
\phi(\hat{k}_1-m)\}{d^3k\over\varepsilon_k} 
\nonumber\\
&&={m\over (2\pi)^3}\int Tr\{\psi^{\dag}(\vec{k},\vec{n}) 
\psi(\vec{k},\vec{n})\}{d^3k\over\varepsilon_k}
={m\over (2\pi)^3}\int (f_1^2+ f_2^2) {d^3k\over\varepsilon_k}=1,
\end{eqnarray}
where we denote $\hat{k}_i=k^\mu_i\gamma_\mu$ and
$\bar{\phi}=\gamma_0 \phi^\dagger \gamma_0$.

The spin structures $S_{i}$ in (\ref{eq1_2}) are constructed in order 
to reproduce equation (\ref{eq0}) without any additional coefficient 
and they are orthonormalized relative to the trace operation in 
(\ref{na5}):
\begin{equation}\label{na5_1}
Tr\{\bar{S}_i(\hat{k}_2+m)S_j(\hat{k}_1-m)\}=\delta_{ij},
\end{equation}
where $\bar{S}_i=\gamma_0 S_i^\dagger \gamma_0$, that is:
\begin{equation}\label{eq1_3} 
\bar{S}_1=-\frac{1}{2\sqrt{2}\varepsilon_k}\gamma_5,\quad
\bar{S}_2=-\frac{\varepsilon_k}{2\sqrt{2} mk\sin\theta}
\gamma_5\left(\frac{2m\hat{\omega}}{\omega\cd p}- 
\frac{m^2}{\varepsilon_k^2}\right).
\end{equation}

We insert expression (\ref{eq1}) for the wave function 
in equation (\ref{eq5d}) and
multiply it on left by $u(k_2)$ and on right by $u(k_1)$.
Using the relation $\sum_{\sigma}u^{\sigma}(k)\bar{u}^{\sigma}(k) =\hat{k}+m$,
we get:
\begin{eqnarray}\label{eq8d} 
&&\left[M^2- 4(\vec{k}\,^2 + m^2)\right] (\hat{k}_2+m)\phi(\hat{k}_1-m) \\
&&=\frac{m^2}{2\pi^3} \int 
\frac{g^2}{4m^2(Q^2+\mu^2)} 
(\hat{k}_2+m) (\hat{k}'_2+m)\phi'(\hat{k}'_1-m)
 (\hat{k}_1-m) \frac{d^3k'}{\varepsilon_{k'}}. 
\nonumber
\end{eqnarray}
Substituting $\phi$ in the form (\ref{eq1_1}), 
multiplying equation (\ref{eq8d}) by $\bar{S}_{i}$ from
(\ref{eq1_3}) and using the orthogonality relation (\ref{na5_1}), we obtain 
the system of two-dimensional integral equations for the components $f_{i}$:
\begin{eqnarray}\label{eq10a}  
\left[M^2- 4(\vec{k}\,^2 + m^2)\right] f_1(k,\theta) &=&
\frac{m^2}{2\pi^3} \int
\left[K_{11}(k,\theta;k',\theta')
f_1(k',\theta')+K_{12}(k,\theta;k',\theta')
f_2(k',\theta')\right]\frac{d^3k'}{\varepsilon_{k'}}
\nonumber\\
\left[M^2- 4(\vec{k}\,^2 + m^2)\right]f_2(k,\theta) &=&
\frac{m^2}{2\pi^3}\int\left[K_{21}(k,\theta;k',\theta')
f_1(k',\theta')+K_{22}(k,\theta;k',\theta')
f_2(k',\theta')\right]\frac{d^3k'}{\varepsilon_{k'}}
\end{eqnarray}
The kernels $K_{ij}$ result from a first integration 
over the azimuthal angle $\varphi'$ of more elementary quantities:
\begin{eqnarray}\label{nz1} 
K_{ij}&=&{1\over m^2\varepsilon_k \varepsilon_{k'}}
\int_0^{2\pi}{\kappa_{ij} \over Q^2+\mu^2}{d\varphi'\over 2\pi}
\end{eqnarray}
with 
\begin{eqnarray}\label{kappa_ij_J0} 
\kappa_{ij}&=&{g^2\over4}\varepsilon_k \varepsilon_k'
Tr\left[\bar{S}_i(\hat{k}_2+m)(\hat{k'}_2+m)S'_j
(\hat{k'}_1-m)(\hat{k}_1-m)\right]\label{nz1_k} 
\end{eqnarray}
However, for notation convenience, we keep in (\ref{eq10a}) the three 
dimensional volume element though the  kernels (\ref{nz1}) are 
$\varphi'$ independent.  The traces (\ref{nz1_k}) are expressed through 
the scalar products of the available four-vectors. The analytical 
expressions for these scalar products and for $\kappa_{ij}$ in the 
variables $k,k',\theta,\theta',\varphi'$ are given in appendix 
\ref{scal_1}.

\section{The $J=1^+$ state}\label{J1}

As mentioned in section \ref{covwf}, 
the two-fermion wave function with $J=1^+$ is determined by six components 
associated to each of the six spin structures one can construct 
\cite{karm81}:
\begin{eqnarray}\label{eq12d}
&&S_{1\mu}=\frac{(k_1- k_2)^{\mu}}{2m^2},\quad
S_{2\mu}=\frac{1}{m}\gamma^{\mu},\quad                           
S_{3\mu}=\frac{\omega^{\mu}}{\omega\cd p},\quad                                       
S_{4\mu}=\frac{(k_1-k_2)^{\mu}\hat{\omega}}{2m\omega\cd p},
\nonumber\\
&&S_{5\mu}=-\frac{i}{m^2\omega\cd p}\gamma_5
\epsilon^{\mu\nu\rho\gamma}k_{1\nu}k_{2\rho}\omega_{\gamma},\quad                              
S_{6\mu}=\frac{m\omega^{\mu}\hat{\omega}}{(\omega\cd p)^2}.
\end{eqnarray}  

The system of equations satisfying by these components can be split in 
two subsystems corresponding respectively to the eigenvalues $a=0,1$ of 
the operator $A^2$ (\ref{ac6}).  The subsystem $a=0$ is -- like in the 
$J=0$ case -- determined by two components whereas the four remaining 
components are related to $a=1$.  The total number of components as 
well as the dimensions of the subsystems (2+4) coincides with the 
results obtained in the standard approach \cite{glazek1}.

\subsection{The case $a=0$}\label{J1_a0}

In this section we consider the state with $a=0$.
We represent the wave function in a form similar to (\ref{eq1}) 
\begin{equation}\label{nz1_1}                                                     
{\mit\Phi}^{\lambda}_{\sigma_2\sigma_1}(k_1,k_2,p,\omega \tau)=
\sqrt{m}e^{\mu}(p,\lambda) 
\bar{u}^{\sigma_2}(k_2)\phi^{(0)}_{\mu}U_c\bar{u}^{\sigma_1}(k_1)\ ,        
\end{equation}    
$e^{\mu}(p,\lambda)$ is the spin-1 polarization vector and  
functions $\phi^{(0)}_{\mu}$ are represented as:
\begin{equation}\label{nz2_0} 
\phi^{(0)}_{\mu}= g^{(0)}_1 S^{(0)}_{1\mu}+ g^{(0)}_2 S^{(0)}_{2\mu}.
\end{equation} 
where $g^{(0)}_i$ are scalar functions and the spin structures $S^{(0)}_{i\mu}$ 
will be expressed in terms of structures (\ref{eq12d}). 

In the reference system where $\vec{k}_1+\vec{k}_2=0$ 
this wave function can be represented in the form:
\begin{equation}\label{nz7}                                                     
{\mit\Phi}^{\lambda}_{\sigma_2\sigma_1}(\vec{k},\hat{n}) = 
\sqrt{m}w^\dagger_{\sigma_2} \psi^{(0)}_{\lambda}(\vec{k},\hat{n})\sigma_yw^\dagger_{\sigma_1}.
\end{equation} 
One can easily check that the function $\vec{\psi}^0(\vec{k},\hat{n})$ --
satisfying  equation (\ref{eqf1}) with $a=0$ --  is proportional to
$\hat{n}$, i.e. it satisfies the condition $\vec{\psi}^0=\hat{n}(\hat{n}\cd \vec{\psi}^0)$. 
Since we are dealing with a $J^{\pi}=1^+$ pseudovector state, 
it has the following general decomposition:
\begin{equation}\label{eq4a} 
\vec{\psi}^0(\vec{k},\vec{n})=\sqrt{\frac{3}{2}}\left\{g^{(0)}_1
\vec{\sigma}\cd\hat{k}
+
g^{(0)}_2 \frac{\vec{\sigma}\cd(\hat{k}\cos\theta-\hat{n})}{\sin\theta}\right\}\hat{n}
\end{equation} 
in which the vector $\hat{n}$ is multiplied by
the two only pseudoscalar structures 
one can construct $\vec{\sigma}\cd\vec{k}$ and $\vec{\sigma}\cd\hat{n}$.

The four-dimensional structures $S^{(0)}_{1,2\mu}$
are built in terms of $S_{i\mu}$ defined in (\ref{eq12d})
in a way to obtain equation (\ref{eq4a}) from (\ref{nz1_1}) written in the c.m.-system.
One finds
\begin{equation}\label{nzf1} 
S^{(0)}_{1\mu}=\frac{\sqrt{3}M}{2\sqrt{2}k}S_{3\mu}\qquad
S^{(0)}_{2\mu}=\frac{\sqrt{3}M}{m\sqrt{2}\sin\theta}
\left(\frac{m^2\cos\theta}{2\varepsilon_k k}S_{3\mu}+S_{6\mu}\right),
\end{equation} 

The normalization condition for the wave function reads:
\begin{eqnarray}\label{na4} 
&&\frac{1}{3}\sum_{\lambda\sigma_2\sigma_1}
\left|{\mit\Phi}^{\lambda}_{\sigma_2\sigma_1}\right|^2=
{m\over (2\pi)^3}\int \Pi^{\mu\nu}Tr\{\phi^{(0)}_\mu(\hat{k}_2+m) 
\phi^{(0)}_\nu(\hat{k}_1-m)\}{d^3k\over\varepsilon_k} 
\\
&=&{m\over 3(2\pi)^3}\int Tr\{\vec{\psi}^{(0)\dag}(\vec{k},\vec{n}) 
\vec{\psi}^{(0)}(\vec{k},\vec{n})\}{d^3k\over\varepsilon_k}
={m\over (2\pi)^3}\int \left[(g^{(0)}_1)^2+ (g^{(0)}_2)^2\right] 
{d^3k\over\varepsilon_k}=1.
\nonumber
\end{eqnarray}
where we have introduced the tensor
\begin{equation}\label{Pi}
\Pi^{\mu\nu}=\frac{1}{3}\sum_{\lambda}e^{\mu *}(p,\lambda)e^{\nu }(p,\lambda)=
\frac{1}{3}\left(\frac{p^{\mu}p^{\nu}}{M^2}-g^{\mu\nu}\right).
\end{equation}

The structures $S^{(0)}_{i}$ are orthonormalized relative to the trace 
operation:
\begin{equation}\label{na5_2}
\Pi^{\mu\nu}Tr\{\bar{S}^{(0)}_{i\mu}(\hat{k}_2+m)
S^{(0)}_{j\nu}(\hat{k}_1-m)\}=\delta_{ij}
\end{equation}
with
$\bar{S}^{(0)}_{i\mu}=\gamma_0S^{(0)\dagger}_{i\mu}\gamma_0=S^{(0)}_{i\mu}$.
The same relation and the orthogonality condition (\ref{na5_2}) hold 
for the structures $S^{(1)}_{i\mu}$ corresponding to $a=1$ and 
constructed in the next section. 

We finally obtain for the components $g^{(0)}_{i}$ a system of two 
equations having the same form than for J=0 -- (\ref{eq10a}) and  
(\ref{nz1}) -- with kernels $\kappa_{ij}$ given by:
\begin{equation}\label{eq10b}
\kappa_{ij}={g^2\over4}\varepsilon_k \varepsilon_k'
\Pi^{\mu\nu} Tr\left[\bar{S}^{(0)}_{i\mu}(\hat{k}_2+m)
(\hat{k'}_2+m)S'^{(0)}_{j\nu}(\hat{k'}_1-m) (\hat{k}_1-m)\right].
\end{equation}

Their explicit analytical expressions are given in appendix \ref{scal_1}.

\subsection{The case $a=1$}\label{J1_a1}

The solution $\vec{\psi}^1(\vec{k},\hat{n})$ corresponding to 
$a=1$ is orthogonal to $\hat{n}$, i.e. it  satisfies $\hat{n}\cd\vec{\psi}^1=0$.
To fulfill this condition, 
it is convenient to introduce the vectors $\hat{k}_\perp$
and $\vec{\sigma}_\perp$ orthogonal to $\hat{n}$:
\[\hat{k}_\perp= \frac{\vec{k}-k\cos\theta\hat{n}}{k\sin\theta}\qquad
\vec{\sigma}_\perp= \vec{\sigma}-(\hat{n}\cd \vec{\sigma})\hat{n}\]

and to write down  $\vec{\psi}^1$ in the form:

\begin{equation}\label{eq4b} 
\vec{\psi}^1(\vec{k},\vec{n})=\frac{\sqrt{3}}{2}\left\{
g^{(1)}_1\vec{\sigma}_\perp
+g^{(1)}_2\left(2\hat{\vec{k}}_\perp 
(\hat{\vec{k}}_\perp \cd \vec{\sigma}_\perp)-\vec{\sigma}_\perp\right)
+g^{(1)}_3 \hat{\vec{k}}_\perp (\vec{\sigma}\cd \vec{n})
+g^{(1)}_4i[\hat{\vec{k}}\times \vec{n}] \right\}
\end{equation}

The four-dimensional representation ${\mit\Phi}^{\lambda}_{\sigma_2\sigma_1}$
 analogous 
to (\ref{nz1_1}), is written in terms of  $\phi^{(1)}_{\mu}$:
\begin{equation}\label{nz2_1} 
\phi^{(1)}_{\mu}= g^{(1)}_1 S^{(1)}_{1\mu}+ g^{(1)}_2 S^{(1)}_{2\mu}+
g^{(1)}_3 S^{(1)}_{3\mu}+ g^{(1)}_4 S^{(1)}_{4\mu}.
\end{equation} 

The four spin structures $S^{(1)}_{j\mu}$ are orthonormalized 
according to (\ref{na5_2}) and read:
\begin{equation}\label{nzf2} 
S^{(1)}_{i\mu}=\sum_j \eta_{ij}S_{j\mu},\quad 
i=1,\ldots,4;\;j=1,\ldots,6, \end{equation} with $S_{j\mu}$ defined in 
(\ref{eq12d}) and the coefficients $\eta_{ij}$ given in appendix 
\ref{app3}.

The normalization condition in terms of $\phi^{(1)}_\mu$ and 
$\vec{\psi}^{(1)}$ exactly coincides with (\ref{na4}) and in terms of the
components $g^{(1)}_i$ is rewritten as:
\begin{equation}\label{nb4} 
{m\over (2\pi)^3}\int \left[(g^{(1)}_1)^2+ (g^{(1)}_2)^2+
(g^{(1)}_3)^2+ (g^{(1)}_4)^2\right] {d^3k\over\varepsilon_k}=1.
\end{equation}

The system of equations for the four scalar functions
$g^{(1)}_{i}$ has also the same form than (\ref{eq10a})
\begin{equation}\label{eq10c}  
\left[M^2-4(\vec{k}\,^2 +m^2)\right] g^{(1)}_i(\vec{k},\hat{n})
=\frac{m^2}{2\pi^3} \int \sum_{j=1}^4 K^{(1)}_{ij}(\vec{k},\vec{k}\,',\hat{n})
g^{(1)}_j(\vec{k}\,',\hat{n})\frac{d^3k'}{\varepsilon_{k'}}
\end{equation}
The kernels $K^{(1)}_{ij}$ are obtained from (\ref{eq10b}) by 
substituting $S^{(1)}$ instead of $S^{(0)}$. Their analytical 
expressions are given in appendix \ref{scal_1}. The details of 
calculation can be found in \cite{These_MMB}.

We are interested here in calculating the mass $M$ of a J=1 state.
The physical solution satisfying the angular condition mentioned above
is given by the superposition of the solutions ($\psi_a,M_a$) with definite $a$ 
\cite{fermions,heidelberg}:
\[ \vec{\psi}=c_0\vec{\psi}^0+c_1\vec{\psi}^1.\]
The corresponding mass is
\[ M^2 = c_0^2 M_0^2+c_1^2 M_1^2\]

\section{Relation with the standard approach}\label{relgl}

In reference \cite{glazek1}, the system of equations
solved for the $J_z=0$ helicity sector had the form:
\begin{eqnarray}\label{eq4gl}
&&\left(M^2-M_0^2\right)\Phi_1(R_{\perp},x) \nonumber\\
&&=\frac{\alpha}{4\pi^2}\int 
\left[V_{11}(R_{\perp},x;R_{\perp}',x')\Phi_1(R_{\perp}',x')
+V_{12}(R_{\perp},x;R_{\perp}',x')\Phi_2(R_{\perp}',x')
\right]R_{\perp}'dR_{\perp}'dx', \nonumber\\
&&\left(M^2-M_0^2\right)\Phi_2(R_{\perp},x)\\
&&=\frac{\alpha}{4\pi^2}\int  
\left[V_{21}(R_{\perp},x;R_{\perp}',x')\Phi_1(R_{\perp}',x')
+V_{22}(R_{\perp},x;R_{\perp}',x')\Phi_2(R_{\perp}',x')\right]
R_{\perp}'dR_{\perp}'dx',\nonumber
\end{eqnarray}
with $M_0^2={R_{\perp}^2+m^2\over x(1-x)}$.
They correspond to equations (3.1a), (3.1b) from \cite{glazek1}
with the notations $k\equiv R_{\perp}$, $q\equiv R'_{\perp}$, $y\equiv x'$.
The coupled wave functions ($\Phi_1,\Phi_2$) should be identified
either to  ($\Phi^{1+},\Phi^{2-}$) or to ($\Phi^{1-},\Phi^{2+}$) sets.
The kernels $V_{ij}$, given by eqs. C1-C4 in \cite{glazek1},
are for the case ($\Phi^{1+},\Phi^{2-}$):
\begin{eqnarray}\label{eqap2}	 
V_{ij}&=&\int_0^{2\pi}\frac{v_{ij}d\varphi'}{x(1-x)x'(1-x')(Q^2+\mu^2)},\nonumber\\
v_{11}&=&R_\perp R_{\perp}'(2xx'-x-x')\nonumber\\
&&+\left[R_{\perp}^2 x'(1-x')+ R'^2_{\perp} 
x(1-x)-m^2(x+x')(2-x-x')\right]\cos\varphi',\nonumber\\ 
v_{12}&=&m \left[R_\perp 
(x+3x'-2xx'-2x'^2)-R'_\perp 
(x'+3x-2x'x-2x^2)\cos\varphi'\right],\nonumber\\ v_{21}&=&m 
\left[R'_\perp (x'+3x-2x'x-2x^2)-R_\perp 
(x+3x'-2xx'-2x'^2)\cos\varphi'\right],\nonumber\\ v_{22}&=&R_\perp 
R_{\perp}'(2xx'-x-x')\cos\varphi'\\ &&+\left[R_{\perp}^2 x'(1-x')+ 
R'^2_{\perp} x(1-x)-m^2(x+x')(2-x-x')\right].\nonumber \end{eqnarray} 
The momentum transfer $Q^2$ reads:
\begin{equation}\label{kernrx}
Q^2= m^2{x'\over x}\left(1-\frac{x}{x'}\right)^2
+\frac{x'}{x}R_{\perp}^2-2R_{\perp}R_{\perp}'\cos\varphi'+\frac{x}{x'}
R_{\perp}'^2+(x'-x)\left(\frac{m^2+R_{\perp}'^2}{x'(1-x')}-M^2\right)
\end{equation}
for $x\leq x'$ and with the replacements $x\leftrightarrow x'$ 
$R_{\perp}\leftrightarrow R_{\perp}'$ for $x\geq x'$.

The functions $\Phi_{i}$ are normalized as:
\begin{equation}\label{eq3a}
\int\left(|\Phi_1(R_\perp,x)|^2 +|\Phi_2(R_\perp,x)|^2\right) 
R_\perp\,dR_\perp\,dx =1.
\end{equation}

We will show that the  equations (\ref{eq10a}) with  kernels  (\ref{eqap1})
corresponding to our $J=0$ state are identical to 
equations (\ref{eq4gl}) with kernels (\ref{eqap2}) for ($\Phi^{1+},\Phi^{2-}$). 
They transform into each other by a
linear combination of the wave functions
and by a  change of variable. The  
relation between the wave functions can be written in the form:
\begin{equation}\label{relwf}
f_j= N(x)\sum_i \Phi_i \;c_{ij}(R_\perp) 
\end{equation}
with a normalization factor 
ensuring the equivalence between the conditions (\ref{na5}) and (\ref{eq3a})
\[N(x)=2^{3/2}\pi m^{-1/2}\sqrt{x(1-x)}\]
and a unitary matrix $\hat{c}$:
\begin{equation}\label{cJ0}
\hat{c}={1\over\sqrt{R_\perp^2+m^2}}
\pmatrix{-R_\perp& m\cr  m &R_\perp}
\end{equation}
Variables $(R_{\perp},x)$ used in (\ref{eq4gl})
 are related to $(k,\theta)$ from (\ref{eq10a}) by:
\begin{equation}\label{eqn11a}
R_{\perp}=k\sin\theta, \qquad
x=\frac{1}{2}\left(1-\frac{k\cos\theta}{\varepsilon_k}\right)
\end{equation}
and their reversed relations:
\begin{equation}\label{eqn11}
k^2=\frac{R_\perp^2+m^2}{4x(1-x)}-m^2,\quad
\cos\theta=\frac{(1-2x)\sqrt{R_\perp^2+m^2}}
{\sqrt{R_\perp^2+m^2(1-2x)^2}}.
\end{equation}

Variables $(\vec{R}_\perp,x)$ can be directly constructed from 
the initial four-momenta as follows.
We introduce the four-vector
$R=k_1-xp$ with $x=\omega\cd k_1/ (\omega\cd p)$ and represent its spatial part as
$\vec{R}=\vec{R}_{\|}+\vec{R_{\perp}}$, where $\vec{R}_{\|}$ is parallel 
to $\vec{\omega}$  and $\vec{R}_{\perp}\cd \vec{\omega}=0$.
Since  by definition  of  $R$,
$R\cd\omega=R_0\omega_0-\vec{R}_{\|}\cd\vec{\omega}=0$, it follows that 
$|R_0|=|\vec{R}_{\|}|$, and hence $\vec{R}^2_{\perp} =-R^2$ is invariant.

For $J=1,a=0$ quantum numbers, corresponding to the state $(1^-,2^+)$ 
from \cite{glazek1}, the relation between the $g^{(0)}_j$ components \
and $\Phi_i$ is the same than (\ref{relwf})
\begin{equation}\label{relwf_1}
g^{(0)}_j = N(x)\sum_i c_{ij}(R_\perp) \Phi_i
\end{equation}
with a $\hat{c}$ matrix
\begin{equation}\label{cJ1a0}
\hat{c}={1\over k\sqrt{R_\perp^2+m^2}}
\pmatrix{R_\perp\varepsilon_k& -k_z m\cr -k_z m &-R_\perp\varepsilon_k}
\end{equation}
Note that in \cite{glazek1} 
$k_z=-k\cos\theta$ with a minus sign due to $\hat{n}=-\hat{z}$).

Inserting (\ref{relwf}) into equations
(\ref{eq10a}), we reproduce  equations (\ref{eq4gl}) 
with kernels $V_{ij}$ which are linear combinations of $K_{ij}$:
\begin{equation}\label{relk}
V_{ij}={N(x')\over N(x)N_1(x')}\sum_{i'j'} 
c_{i'i}(R_{\perp}) K_{i'j'} c_{j'j}(R'_{\perp}),
\qquad N_1(x')={\alpha\pi\over m^2} x'(1-x'). 
\end{equation}
Substituting here the kernels $K_{ij}$ from (\ref{nz1})  for $J=0$, we get
the expressions for $V_{ij}$ given by  (\ref{eqap2}). A similar identity is
obtained for the kernels $K_{ij}$ of $J=1,a=0$ state (\ref{eq11})  and the
kernels $V_{ij}$ for the corresponding ($1^-,2^+$) state considered in 
\cite{glazek1}. 
Hence, both systems of equations are strictly equivalent.

It is useful to explain the origin of the linear transformations
(\ref{relwf}) and (\ref{relwf_1}). 
They result from using different representations of the 
spinors and different normalizations of the wave functions.
Namely, instead of eq. (\ref{eq74}) the following spinor is defined in 
\cite{glazek1}:
\begin{equation}\label{eqn2}
u^{LF}_{\sigma}(k)=\sqrt{\frac{2}{k_0+k_3}}\left[m\Lambda^- 
+(k_0+k_3 +\gamma_0\vec{\gamma}^\perp\cd\vec{k}^\perp)\Lambda^+\right]
\left(\begin{array}{l}
w_\sigma\\0\end{array}\right)
\end{equation}
where
\[
\Lambda^\pm=\frac{1}{4}\gamma^{\mp}\gamma^{\pm},
\quad \gamma^{\pm}=\gamma_0\pm\gamma_3, \quad \vec{k}^\perp=(k_1,k_2).\]
It can be transformed to the form:
\begin{equation}\label{eqn3}
u^{LF}_{\sigma}(k)=\sqrt{\frac{1}{2(k_0+k_3)}}
\left(
\begin{array}{l}
m+k_0+\vec{\sigma}\cd\vec{k}\,\sigma_3\\
(k_0 - m)\,\sigma_3+\vec{\sigma}\cd\vec{k}
\end{array}
\right)w_\sigma
\end{equation}
In order to relate the two wave functions, one may express the spinors
$u^{LF}$ in terms of $u$:
\begin{equation}\label{eqn4}
u^{LF}_{\sigma}(k)=\sum_{\sigma'=\pm\frac{1}{2}}
\beta_{\sigma\sigma'}(k)u_{\sigma'}(k).
\end{equation}
With the explicit formulas (\ref{eq74}) and (\ref{eqn3}) we find:
\begin{equation}\label{eqn5}
\beta_{\sigma\sigma'}(k)=
\frac{1}{2m}\bar{u}_{\sigma'}(k)u^{LF}_{\sigma}(k)=
\sqrt{\frac{k_0+m}{2(k_0+k_3)}}w^{\dagger}_{\sigma'}
\left(1+\frac{\vec{\sigma}\cd\vec{k}\,\sigma_3}{k_0+m}\right) w_{\sigma}
\end{equation}
Note the opposite order of indices in both sides
of (\ref{eqn5}). Below we give explicitly these matrix elements
in the c.m.-system $\vec{k}_1+\vec{k}_2=0$:
\begin{eqnarray}\label{eqn6}
&&\beta_{\frac{1}{2}\frac{1}{2}}(k_1)=\beta_{-\frac{1}{2}-\frac{1}{2}}(k_1)=
\frac{(k^+ +m)}{\sqrt{2k^+(k_0+m)}},
\quad
\beta_{\frac{1}{2}-\frac{1}{2}}(k_1)=
-\beta^*_{-\frac{1}{2}\frac{1}{2}}(k_1)=
\frac{k_\perp e^{i\phi}}{\sqrt{2k^+(k_0+m)}},
\nonumber\\
&&\beta_{\frac{1}{2}\frac{1}{2}}(k_2)=
\beta_{-\frac{1}{2}-\frac{1}{2}}(k_2)=
\frac{(k^- +m)}{\sqrt{2k^-(k_0+m)}},
\quad
\beta_{\frac{1}{2}-\frac{1}{2}}(k_2)=
-\beta^*_{-\frac{1}{2}\frac{1}{2}}(k_2)=
\frac{-k_\perp e^{i\phi}}{\sqrt{2k^-(k_0+m)}},
\nonumber\\
&&
\end{eqnarray}
with $k_1+ik_2=k_{\perp}e^{i\phi}$.
Coefficients  $\beta(k_1)$  relate the 
spinors of particle 1 with momentum $\vec{k}_1=\vec{k}$, whereas 
coefficients  $\beta(k_2)$  relate the spinors of particle 2 
with momentum $\vec{k}_2=-\vec{k}$). 
For the conjugated spinors $\bar{u}$ we get:
\begin{equation}\label{eqn6a}
\bar{u}^{LF}_{\sigma_1}(k_1)=\sum_{\sigma'_1=\pm\frac{1}{2}}
\beta^*_{\sigma_1\sigma'_1}(k_1)\bar{u}_{\sigma'_1}(k_1),
\quad
\bar{u}^{LF}_{\sigma_2}(k_2)=\sum_{\sigma'_2=\pm\frac{1}{2}}
\beta^*_{\sigma_2\sigma'_2}(k_2)\bar{u}_{\sigma'_2}(k_2),
\end{equation}
The matrix $\beta$ is unitary: 
$\sum_{\sigma'}\beta^\dagger_{\sigma\sigma'}\beta_{\sigma'\sigma''}=
\sum_{\sigma'}\beta^*_{\sigma'\sigma}\beta_{\sigma'\sigma''}=
\delta_{\sigma\sigma''}$. 
Substituting the spinors 
(\ref{eqn6a}) into the wave function $\Psi$ defined by means of the spinors
$\bar{u}^{LF}$, we get the relation:
\begin{equation}\label{eqn7}
\Psi^{\sigma_1\sigma_2}=\sum_{\sigma'_1\sigma'_2} 
\beta^*_{\sigma_1\sigma'_1}(k_1) \beta^*_{\sigma_2\sigma'_2}(k_2)
\psi_{\sigma'_2\sigma'_1}
\end{equation}
Note the opposite order of indices at $\Psi^{\sigma_1\sigma_2}$ and 
$\psi_{\sigma'_2\sigma'_1}$, in accordance with the above definitions.
The matrix elements of the wave function $\psi(\vec{k},\vec{n})$ for 
$J=0$, eq. (\ref{eq0}), are the following:
\begin{equation}\label{eqn7J0}
\psi_{\frac{1}{2}\frac{1}{2}}= 
\psi^*_{-\frac{1}{2}-\frac{1}{2}}= \frac{i}{\sqrt{2}}f_2 e^{-i\phi},
\quad
\psi_{\frac{1}{2}-\frac{1}{2}}=
-\psi_{\frac{1}{2}-\frac{1}{2}}= -\frac{i}{\sqrt{2}}f_1.
\end{equation}

We substitute  in  (\ref{eqn7}) the coefficients (\ref{eqn6}) and the
wave function (\ref{eqn7J0}), following the paper \cite{glazek1}, separate the phase factor:
\[\Psi^{\sigma_1\sigma_2}(\vec{R}^\perp,x)=\sum_{m} \frac{e^{i m \phi}}{\sqrt{2\pi}}
\Phi^{\sigma_1\sigma_2}(R,x,m)\]
and  introduce the following linear combinations:
\begin{eqnarray}\label{eq2a}
\Phi^{1\pm}(R,x)&=&\frac{1}{\sqrt{2}}[\Phi^{\frac{1}{2}\frac{1}{2}}(R,x;-1)\pm
\Phi^{-\frac{1}{2}-\frac{1}{2}}(R,x;1)],
\nonumber\\
\Phi^{2\pm}(R,x)&=&\frac{1}{\sqrt{2}}[\Phi^{\frac{1}{2}-\frac{1}{2}}(R,x;0)\pm
\Phi^{-\frac{1}{2}\frac{1}{2}}(R,x;0)]
\end{eqnarray}
normalized according to (\ref{eq3a}).
In this way we reproduce for $\Phi^{1+},\Phi^{2-}$
the relation (\ref{relwf}) with 
coefficients (\ref{cJ0}). For the other pair of
components we get $\Phi^{1-}=\Phi^{2+}=0$.

Similarly, taking the matrix elements of the 
wave function $\vec{\psi}^{(0)}(\vec{k},\hat{n})$ for 
$J=1,a=0$, equation (\ref{eq4a}), projected on the direction $\hat{n}$
(we remind that in the standard approach $\hat{n}||-z$), we find:
\begin{eqnarray*}
&&\vec{n}\cd\vec{\psi}^{(0)}_{\frac{1}{2}\frac{1}{2}}= 
\vec{n}\cd\vec{\psi}^{(0)*}_{-\frac{1}{2}-\frac{1}{2}}= 
\frac{i}{\sqrt{2}k}(g^{(0)}_1R_\perp -g^{(0)}_2 k_3)e^{-i\phi},
\\
&&\vec{n}\cd\vec{\psi}^{(0)}_{\frac{1}{2}-\frac{1}{2}}=
\vec{n}\cd\vec{\psi}^{(0)}_{\frac{1}{2}-\frac{1}{2}}= 
-\frac{i}{\sqrt{2}k}(g^{(0)}_1 k_3+g^{(0)}_2 R_\perp).
\end{eqnarray*}
In this way we reproduce for $(\Phi^{1-},\Phi^{2+})$ the relation 
(\ref{relwf_1}) with  
coefficients (\ref{cJ1a0}), whereas we get $\Phi^{1+}=\Phi^{2-}=0.$

\section{Asymptotical properties and
critical coupling constant}\label{asympt}

The stability of solutions is determined by the asymptotical behavior 
of kernels $K_{ij}$.  This can be considered either for a fixed $k(k')$ 
in the limit $k'(k)\rightarrow\infty$ or in the limit of both 
$k,k'\rightarrow\infty$ with a fixed ratio $\gamma={k'\over k}$.  We 
illustrate in what follows the analyzing method for the $J=0$ state.  
The relevant results for $J=1$ are summarized at the end of the 
section.

For a fixed value of $k(k')$ we get:
\begin{eqnarray}\label{as1}
&&K_{11}(k,\theta,k',\theta')\sim\left\{
\begin{array}{ll}
\frac{c_{11}}{k}  &\mbox{if $k\to\infty$,  $k'$ fixed}\\&\\
\frac{c'_{11}}{k'} &\mbox{if $k'\to\infty$, $k$ fixed}
\end{array}\right.\nonumber\\
&&K_{12}(k,\theta,k',\theta')\sim\left\{
\begin{array}{ll}
\frac{c_{12}}{k}&\mbox{if $k\to\infty$,  $k'$ fixed}\\&\\
c'_{12}     &\mbox{if $k'\to\infty$, $k$ fixed}
\end{array}\right.\nonumber\\
&&K_{21}(k,\theta,k',\theta')\sim\left\{
\begin{array}{ll}
c_{21}       &\mbox{if $k\to\infty$, $k'$ fixed} \\&\\
\frac{c'_{21}}{k'} &\mbox{if $k'\to\infty$, $k$ fixed}
\end{array}\right.\nonumber\\
&&K_{22}(k,\theta,k',\theta')\sim\left\{
\begin{array}{ll}
c_{22} &\mbox{if $k\to\infty$, $k'$ fixed}\\&\\
c'_{22}&\mbox{if $k'\to\infty$, $k$ fixed}
\end{array}\right.
\end{eqnarray}
with coefficients $c_{11},c_{12},c_{21},c_{22}$ depending  on $k',\theta,\theta'$ 
and $c'_{11},c'_{12},c'_{21},c'_{22}$ depending on $k,\theta,\theta'$.

One has for instance
\begin{equation}\label{as7}
c_{22}=\frac{\alpha\pi k'\sin\theta\sin\theta'}
{m^2(\varepsilon_{k'}-k'\cos\theta\cos\theta'
+|\varepsilon_{k'}\cos\theta-k'\cos\theta'|)} >0
\end{equation}
and $c'_{22}$ obtained form $c_{22}$ by the replacement
$k'\rightarrow k,\theta\leftrightarrow \theta'$.
Note that these behaviors are obtained once the integration over 
$\varphi'$ in (\ref{nz1}) is performed.

It follows from (\ref{as1})
that the second iteration of  $K_{11}$ converges at $k'\to\infty$:
\[\int K_{11}G_0K_{11}\frac{d^3k'}{\varepsilon_{k'}}\sim
\int \frac{1}{k'}\frac{1}{k'^2}\frac{1}{k'}\frac{k'^2dk'}{k'}=
\int^L \frac{dk'}{k'^3}\propto const. \]
where $G_0\sim 1/k'^2$ is the intermediate propagator.
The integrals 
$\int K_{21}G_0K_{11}{d^3k'\over\varepsilon_{k'}}$,
$\int K_{11}G_0K_{12}{d^3k'\over\varepsilon_{k'}}$ and 
$\int K_{21}G_0K_{12}{d^3k'\over\varepsilon_{k'}}$ also converge 
whereas the second iteration of $K_{22}$ is logarithmically divergent:
\[\int K_{22}G_0K_{22}\frac{d^3k'}{\varepsilon_{k'}}\sim
\int^L \,const \,\frac{1}{k'^2} \frac{k'^2dk'}{k'}=
\int^L\frac{dk'}{k'}\sim \log(L).\]

The integral $\int K_{12}G_0K_{22}{d^3k'\over\varepsilon_{k'}}$,
$\int K_{22}G_0K_{21}{d^3k'\over\varepsilon_{k'}}$ and
$\int K_{12}G_0K_{21}{d^3k'\over\varepsilon_{k'}}$
are also logarithmically divergent,  
as a manifestation of the logarithmic divergence in the LFD box fermion diagram.

In the domain where both $k,k'$ tend to infinity with a fixed ratio ${k'\over k}=\gamma$,
it is useful to introduce the functions
\begin{equation}\label{eqn16}
K_{ii}=-\frac{\pi\alpha}{m^2}\left\{
\begin{array}{ll}
\sqrt{\gamma}      \;A_{ii}(\theta,\theta',\gamma)   & \mbox{if $\gamma\leq 1$}\\
{1\over\sqrt\gamma}\;A_{ii}(\theta,\theta',1/\gamma) & \mbox{if $\gamma\geq 1$}
\end{array}\right.
\end{equation}
where we have extracted for convenience the factor $\sqrt{\gamma}$. 
We find for $A_{11}$:
\begin{equation}\label{eq14a}
A_{11}(\theta,\theta',\gamma)=
\frac{1}{\sqrt{\gamma}}\int_0^{2\pi}\frac{d\varphi'}{2\pi}
\frac{2\gamma(1-\cos\theta\cos\theta')-
(1+\gamma^2)\sin\theta\sin\theta'\cos\varphi'}
{(1+\gamma^2)(1+|\cos\theta-\cos\theta'|-\cos\theta\cos\theta')
-2\gamma\sin\theta\sin\theta'\cos\varphi'},
\end{equation}
and for   $A_{22}$:
\begin{equation}\label{eq14a_1}
A_{22}(\theta,\theta',\gamma)=
-\frac{1}{\sqrt{\gamma}}\int_0^{2\pi}\frac{d\varphi'}{2\pi}
\frac{(1+\gamma^2)\sin\theta\sin\theta'-
2\gamma(1-\cos\theta\cos\theta')\cos\varphi'}
{(1+\gamma^2)(1+|\cos\theta-\cos\theta'|-\cos\theta\cos\theta')
-2\gamma\sin\theta\sin\theta'\cos\varphi'}.
\end{equation}
with $A_{11}(\theta,\theta',\gamma)\sim +\sqrt{\gamma}$ and 
$A_{22}(\theta,\theta',\gamma)\sim-1/\sqrt{\gamma}$ for $\gamma\to0$.

Comparing the above formulas, we see that the dominating kernel is 
$K_{22}$. It does not decrease in any direction of the $(k,k')$ plane, 
whereas in the domain $k\to\infty$, $k'$ fixed (and vice versa) 
$K_{11}$ decreases.  However, in the domain ${k'\over k}=\gamma$ fixed, 
none of the diagonal kernels decreases.  The positive function 
(\ref{eq14a}) corresponds to an attractive kernel $K_{11}$ whereas the 
unbounded function (\ref{eq14a_1}) correspond to a repulsive $K_{22}$ 
(note that the relation (\ref{eqn16}) between $K_{ii}$ and $A_{ii}$ 
contains the sign minus).

The preceding results will be used to find the critical value of the 
coupling constant, above which the solutions are not stable as a 
function of the cut-off value $k_{max}$.  Since  $K_{22}$ is repulsive 
it cannot generate  a collapse.  The formalism is therefore developed 
in the one channel problem, e.g. the $f_1$ component with the kernel 
$K_{11}$ satisfying the equation
\begin{equation}\label{as2_0}  
\left[M^2-4(k^2+m^2)\right] f(k,z) =\frac{m^2}{2\pi^3} \int
K(k,z;k',z')f(k',z')\frac{d^3k'}{\varepsilon_{k'}}.
\end{equation}
where $z=\cos\theta$ and the channel indices are hereafter omitted.

Our further analysis leans on a method developed by Smirnov
\cite{smirnov} and uses the fact that at $k\to\infty$ the integral in 
r.h.-side of (\ref{as2_0}) is dominated by $k'\propto k$. Indeed,
when $k'=fixed$, $k\to\infty$, the kernel $K=K_{11}$ in 
(\ref{as2_0}) decreases like $1/k$ (see (\ref{as1})), that results in 
the $1/k^3$ asymptotics for $f$. We will see below that when $k'\propto
k$, the wave function decreases more slowly than $1/k^3$:
\begin{equation}\label{pl}
f(k,z) \sim {h(z)\over k^{2+\beta}},  \qquad 0\le\beta< 1. 
\end{equation}
Therefore, the integration domain $k'\propto k$ gives  dominating 
contribution. Hence, in the integrand of eq.  (\ref{as2_0}) one can 
replace the kernel by its asymptotics (\ref{eqn16}) and substitute the 
asymptotical form of the wave function (\ref{pl}).

Let us put $k'=\gamma k$ and consider the  limit $k\to\infty$
in which equation tends to
\[-4 f(k,z)=\frac{m^2}{\pi^2} 
\int_0^{\infty} \gamma d\gamma \int_{-1}^{+1}dz' 
K(k,z;\gamma k,z')f(\gamma k,z') \]
where we have neglected the binding energy, supposing that it is finite.

The integral over $\gamma$ can be split in two terms in order
to use the limiting values of the kernel in its form (\ref{eqn16})
\begin{eqnarray*}
4f(k,z) &=& {\alpha\over\pi} \int_{-1}^1 dz' 
\left\{ \int_0^1        \gamma d\gamma \;\sqrt{\gamma}        \; A(\theta,\theta',\gamma) f(\gamma k,z') 
       +\int_1^{\infty} \gamma d\gamma \;{1\over\sqrt{\gamma}}\; A(\theta,\theta',1/\gamma) f(\gamma k,z')\right\} 
\end{eqnarray*}
Both terms can be gathered by putting $\gamma'={1\over\gamma}$ in the second one 
\begin{equation}\label{as2_1}  
4f(k,z) ={\alpha\over\pi} \int_{-1}^1 dz' 
 \int_0^1  d\gamma\; A(\theta,\theta',\gamma)
 \left\{\gamma^{3/2} f(\gamma k,z')+\gamma^{-5/2} f({k\over\gamma},z')\right\} 
\end{equation}
Inserting in (\ref{as2_1}) the wave function in the asymptotical form
(\ref{pl}) leads to the eigenvalue equation
\begin{equation}\label{gHg}
h(z)=\alpha\int_{-1}^{+1} dz' H_{\beta}(z,z')\;h(z')  
\end{equation}
with
\begin{equation}\label{H_beta}
H_{\beta}(z,z')=\int_0^1{1\over2\pi\sqrt\gamma}d\gamma\; 
A(z,z',\gamma)\;\cosh{(\beta\log\gamma)} 
\end{equation} 
The relation 
between the coupling constant $\alpha$ and the coefficient $\beta$, 
determining the power law  of the asymptotic wave function, can be found in 
practice by solving the eigenvalue equation (\ref{Sta_5}) for a fixed value of $\beta$
\begin{equation}\label{Sta_5}
\lambda_{\beta} \, h(z)=\int_{-1}^{+1} dz' H_{\beta}(z,z')\;h(z')  
\end{equation}
and taking
\begin{equation}\label{a_b}
 \alpha(\beta) ={1\over\lambda_{\beta}} 
 \end{equation}
One should notice that the contribution $\gamma\to 0$ in the first term 
of integral (\ref{as2_1}) introduces corrections of higher order than 
${1\over k^{2+\beta}}$ and does not modify the asymptotics of the 
solution $f$. 

The kernel $A(z,z',\gamma)$ in eq. (\ref{H_beta}) is positive (see eq.  
(\ref{eq14a})).  The function $\cosh{(\beta\log\gamma)}$ and, hence, 
the kernel $H_{\beta}(z,z')$ have minimum at $\beta=0$.  This value of 
beta corresponds to maximal -- critical -- value of $\alpha$ 
($\alpha_c$).

It is worth noticing that if $A$ does not depend on $z,z'$, 
$A(\gamma,z,z')=A(\gamma)$, one gets \cite{smirnov}:  
\begin{equation}\label{cut4} {1\over\alpha_c}=\int_0^1 d\gamma \; 
{A(\gamma)\over\pi\sqrt\gamma} \end{equation} Applying this condition 
to the non relativistic Schr\"odinger equation with potential 
\[V(r)=-{\alpha'\over r^2}=-{\alpha\over2m\pi}{1\over r^2},\] 
that corresponds to
$A(\gamma)=1$, one gets the critical coupling constant 
$\alpha_c={\pi\over2}$, that is $\alpha'_c={1\over4m}$ in agreement 
with \cite{ll}.  For the LFD Yukawa model, $A_{11}$ is majorated by 
$\sqrt{\gamma}$.  Inserting this value in (\ref{cut4}), we conclude 
that $\alpha_c>\pi$, in coincidence with estimation given in 
\cite{mck_prd}.

Let us now examine the asymptotics of component $f_2$.  We consider 
equation (\ref{eq10a}) on a finite interval $0\leq k \leq k_{max}$ and 
investigate the behavior of $f_2$ at $k\gg m$.  Since kernel $K_{22}$, 
in contrast to $K_{11}$, tends to a constant at $k\to\infty$ its 
contribution leads to the $1/k^2$ asymptotics for $f_2$:
\begin{equation}\label{f2}
f_2(k,z)\propto \frac{h_2(k_{max},z)}{k^2}.
\end{equation}

It seems at first glance that
with $1/k^2$ asymptotics one gets a logarithmic divergence 
of the integral in the r.h.s. of equation (\ref{eq10a}) for $f_2$:
\[\int K_{22}f_2\frac{d^3k'}{\varepsilon_{k'}} \sim
\int_0^{k_{max}} \frac{1}{k'^2}\frac{k'^2dk'}{k'}\sim \log(k_{max})\].
 However from the asymptotics of equation (\ref{eq10a})
it follows that the asymptotical coefficient $h_2(k_{max},z)$ decreases 
as \[ h_2(k_{max},z)=\frac{a}{1+b\log(k_{max})} \] thus ensuring the 
convergence of the r.h.s. integral in (\ref{eq10a}).

For $J=1,a=0$ state, the  $A_{22}$ function
behaves as $\sim+{1\over\sqrt\gamma}$ when $\gamma\to 0$,
what corresponds to a singular attractive $K_{22}$ kernel
and generates a collapse.

A similar situation occurs for $J=1,a=1$. Inspection of 
the four diagonal kernels shows that at $\gamma\to 0$,
function $A_{22}$ behaves also as $A_{22}\sim +{1\over\sqrt{\gamma}}$,
namely:
\[
A^{(j=1,a=1)}_{22}(\theta,\theta',\gamma\to 0)
=\frac{1}{\sqrt{\gamma}}\frac{\sin^2\theta \sin^2\theta'}
{4(1+|\cos\theta-\cos\theta'|-\cos\theta\cos\theta')}.\]
This singular attraction is responsible for the $J=1,a=1$ instability.

\section{Results}\label{num}

Let us first present the results given by the single equation for $f_1$ with 
kernel $K_{11}$ in the $J=0$ case. 
In all the calculations, the constituent masses were taken equal to $m$=1 
and the mass of the exchanged scalar is $\mu$=0.25.

\begin{figure}[h]
\begin{center}
\epsfxsize=10.0cm\mbox{\epsffile{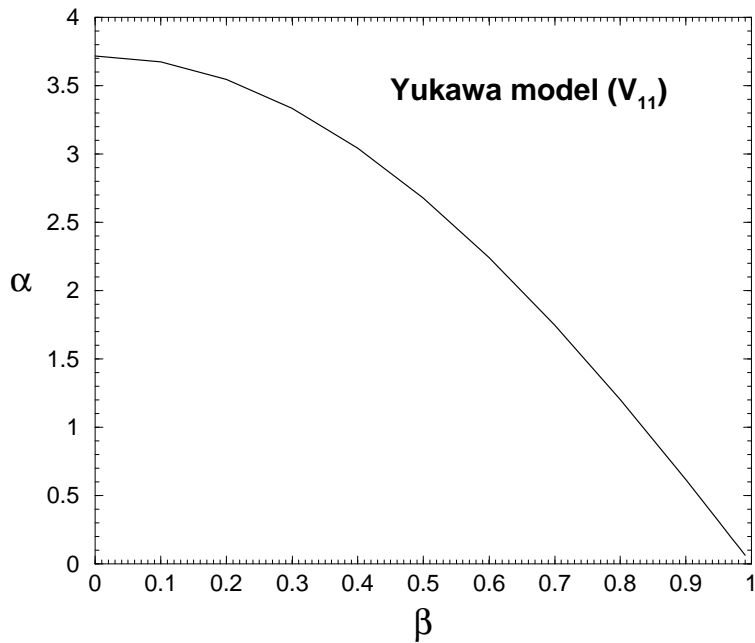}}
\end{center}
\caption{Function $\alpha(\beta)$ for LFD Yukawa $K_{11}$.}
\label{alpha_beta} 
\end{figure}

The numerical solution of equation (\ref{Sta_5}) with the function 
$A_{11}(\gamma,z,z')$ given by (\ref{eqn16}) is plotted in Figure 
\ref{alpha_beta}.  

The critical coupling constant is obtained for 
$\beta=0$ for which the eigenvalue is $\lambda_c=0.269$.  It 
corresponds, according to (\ref{a_b}), to $\alpha_c=3.72$, in agreement 
with our numerical estimations $\alpha_c>\pi$ \cite{mck_prd}.

\begin{figure}[h]
\begin{center}
\epsfxsize=10.cm\mbox{\epsffile{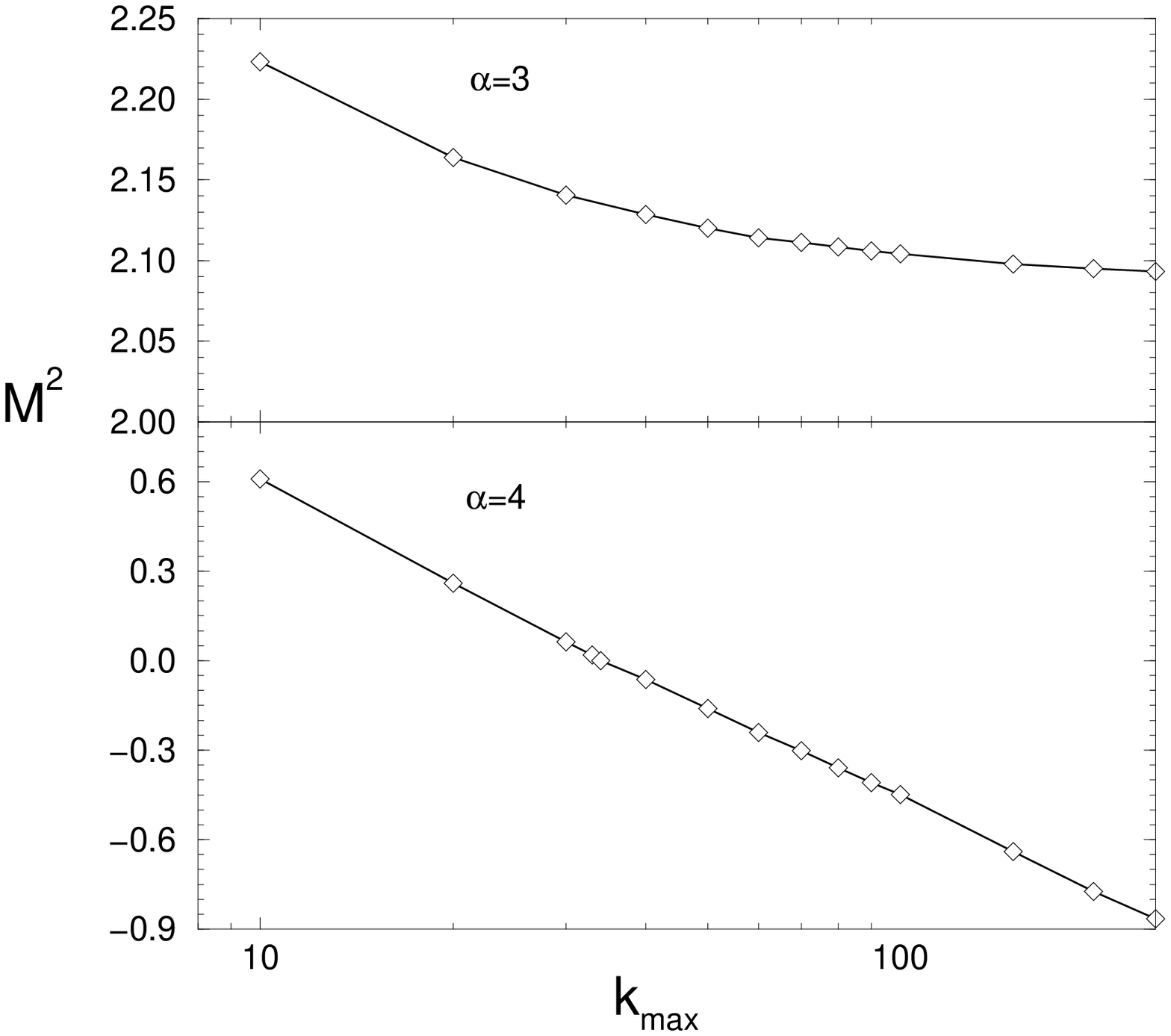}}
\end{center}
\caption{Cutoff dependence of the binding energy in the $J=0$ or $(1+,2-)$
state,
in the one-channel problem ($f_1$), for two fixed values of the coupling 
constant below and above the critical value.}\label{B_kmax}
\end{figure}

We have plotted in Figure \ref{B_kmax} the
mass square $M^2$ of the two fermion system  as a function of the cutoff
$k_{max}$ for two fixed values of the coupling constant below and above  the
critical value $\alpha_c=3.72$. 
In our calculations the cutoff appears directly as the maximum
value $k_{max}$  up to which the integrals in (\ref{eq10a}) are performed. One
can see two dramatically different behaviors depending on the value of the
coupling constant $\alpha$. For $\alpha=3$, the
result is convergent. On the contrary, for $\alpha=4$, i.e. $\alpha>\alpha_{c}$, 
the result is clearly divergent: $M^2$ decreases logarithmically as a function of $k_{max}$
and becomes even negative. 
This divergence is due only to the large $k$ behavior of
$K_{11}$.  Though the negative values of $M^2$ are physically meaningless, they
are formally  allowed by the equations (\ref{eq10a}) and (\ref{eq4gl}). 
The first
degree of $M$ does not enter neither in the equation nor in the kernel, and
$M^2$  crosses zero without any singularity.  The value of the critical
$\alpha$ does not depend on the exchange mass $\mu$. For $\mu\ll m$, e.g.
$\mu\approx0.25$, its existence is not relevant in describing physical states  
since any solution with positive $M^2$, stable relative to cutoff, corresponds
to $\alpha<\alpha_c$. For $\mu\sim m$ one can reach the critical $\alpha$  for
positive, though small values of $M^2$.

Let us consider now the two-channel problem.  The kernel dominating in  
asymptotics is $K_{22}$.  In the case $J=0$ it is positive and 
corresponds to repulsion. Because of that, this kernel does not 
generate by itself any instability but cannot prevent from the collapse 
in the first channel (for enough large $\alpha$), since due to coupling 
between the two  channels the singular potential in channel 1 "pumps 
out" the wave function from channel 2. We would like to emphasize that 
the divergence in the $J=0$ case, when it happens, is not associated 
with the non decreasing behavior of the $K_{22}$ kernel but with the 
existence of a critical value of the coupling constant separating two 
dynamical regimes. This property is due only to the attraction and 
large $k$ behavior of $K_{11}$.  

\begin{figure}[h]
\begin{center}
\mbox{\epsfxsize=10.cm\epsfysize=10.cm\epsffile{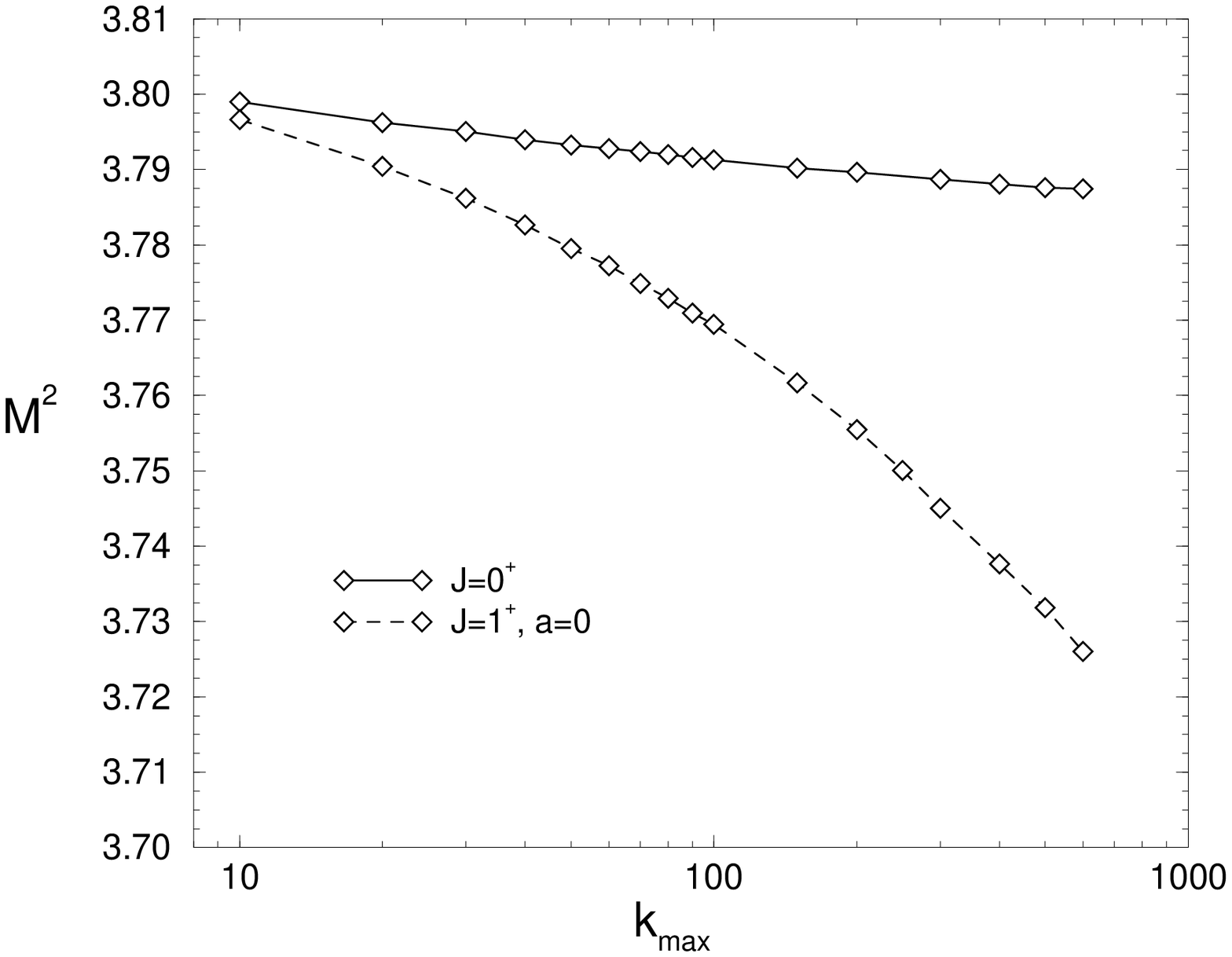}}
\end{center}
\caption{Cutoff dependence of  $M^2$ for $J=0$  $(1+,2-)$ and 
$J=1,a=0$ $(1-,2+)$ states, in full (two-channel) problem, for $\alpha=1.184$.}
\label{M2_kmax}
\end{figure}

\begin{figure}[h] 
\begin{minipage}[t]{80mm}
\begin{center}
\mbox{\epsfxsize=8.0cm\epsfysize=8.0cm\epsffile{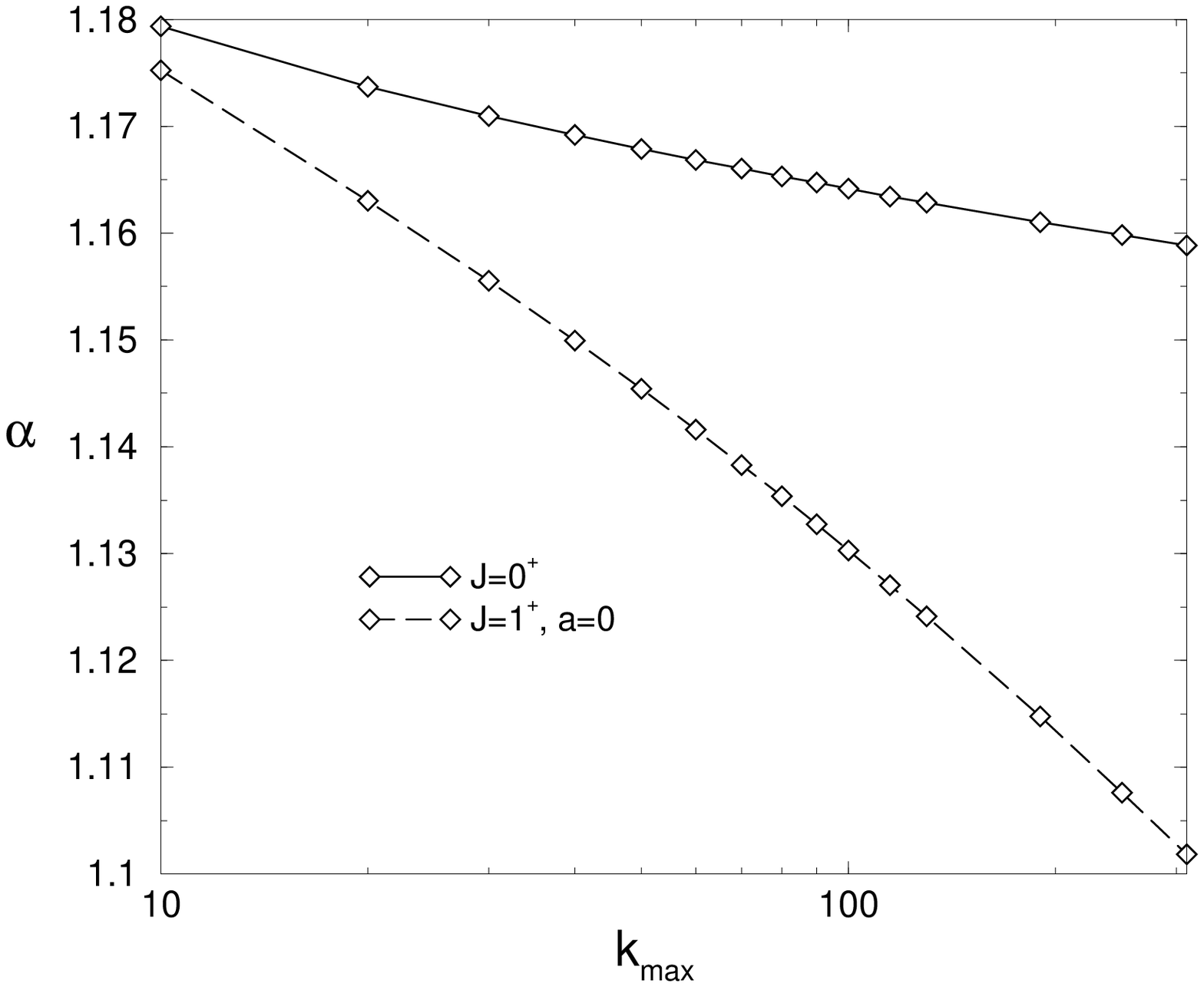}}
\end{center}
\caption{Cutoff dependence of the coupling constant, for $J=0$  and 
$J=1,a=0$  states, in full (two-channel) problem, for 
$B=0.05$.} \label{alpha_kmax_50} \end{minipage}\hspace{0.5cm} 
\begin{minipage}[t]{80mm}
\begin{center}
\mbox{\epsfxsize=8.0cm\epsfysize=8.0cm\epsffile{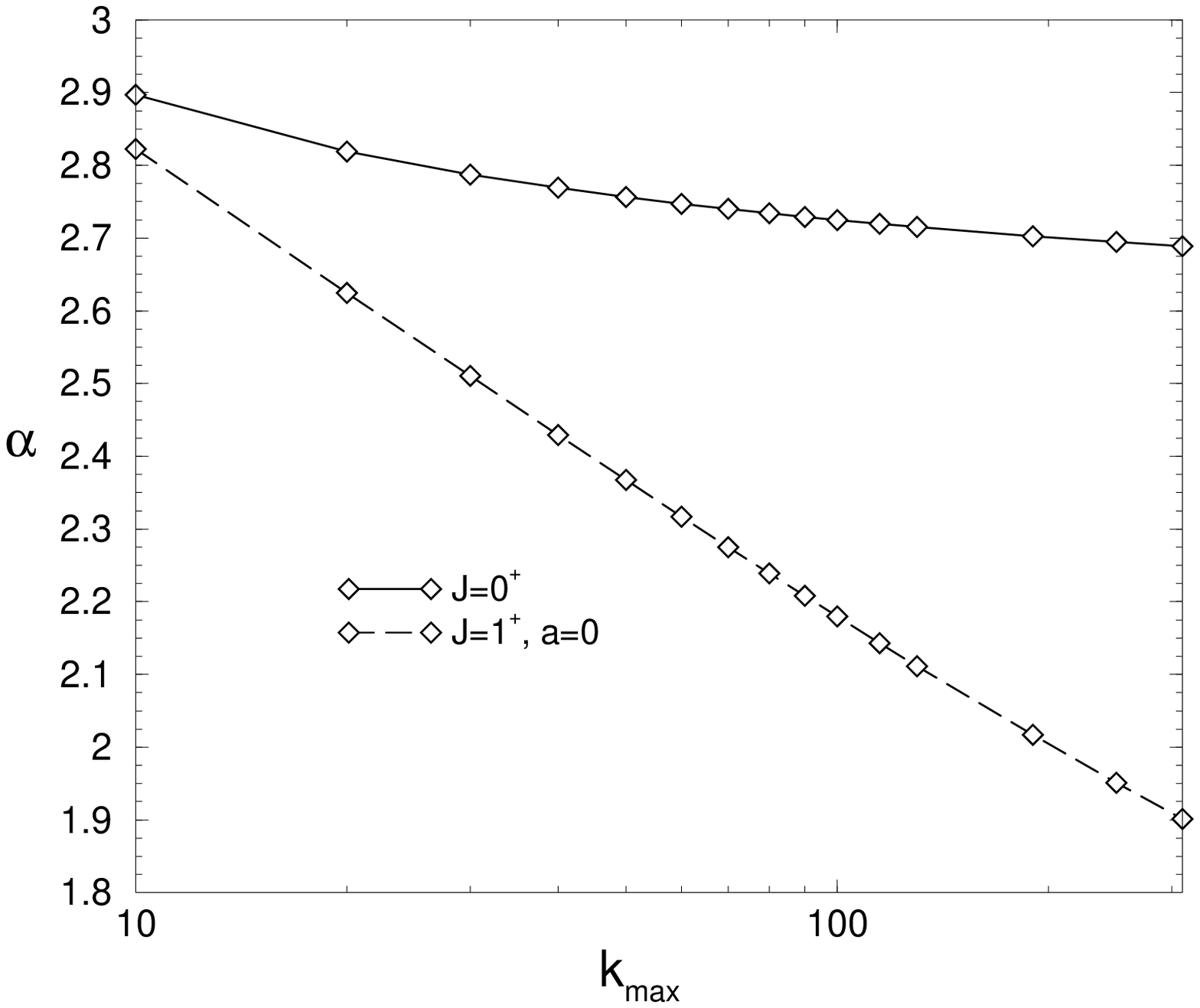}}
\end{center}
\caption{Cutoff dependence of the coupling constant, for $J=0$  and 
$J=1,a=0$  states, in full (two-channel) problem, for $B=0.5$.}
\label{alpha_kmax_500}
\end{minipage}
\end{figure}

In the coupled equations system 
(\ref{eq10a}) the situation with the cutoff dependence is thus the same 
as for one channel.  In Figure \ref{M2_kmax} is displayed the variation 
of $M^2$ for $J=0$ -- or $(1+,2-)$ and $J=1,a=0$ -- or $(1-,2+)$ states 
as a function of the cutoff $k_{max}$. The value of the coupling 
constant is $\alpha=1.184$, the same that in Figure 2 of 
\cite{glazek1}, below the critical value. Our numerical results are in 
agreement with those presented in this figure for a cutoff $\Lambda 
\leq 100$, but our calculation at larger $k_{max}$ leads to different 
conclusion for the $J=0$ state. One remarks a qualitatively different 
behavior of the two states. In what concerns  $J=0$, the numerical 
results become flatter when $k_{max}$ increases, with less than a 0.5\% 
variation in $M^2$ when changing $k_{max}$ between $k_{max}$=10 and 
300.  The same kind of behavior is manifested when studying the cutoff 
dependence of the coupling constant for a fixed value of the binding 
energy. Figures \ref{alpha_kmax_50} and \ref{alpha_kmax_500} show the 
$\alpha(k_{max})$ variation for B=0.05 and B=0.5.  
The J=0 state is very well fitted by a 
law $\alpha(k_{max})=\alpha_{\infty}+{\alpha_0\over k_{max}^{\nu}}$ 
with parameters
$\alpha_{\infty}=1.140$, $\alpha_{0}=0.0639$ and $\nu=0.210$ for B=0.05
and 
$\alpha_{\infty}=2.648$, $\alpha_{0}=0.809$ and $\nu=0.514$ for B=0.50.
We thus conclude to the stability of the state with $J=0$, as expected from our
analysis in section \ref{asympt}.

\begin{figure}[h]
\begin{minipage}[t]{80mm}
\begin{center}\epsfxsize=80mm\mbox{\epsffile{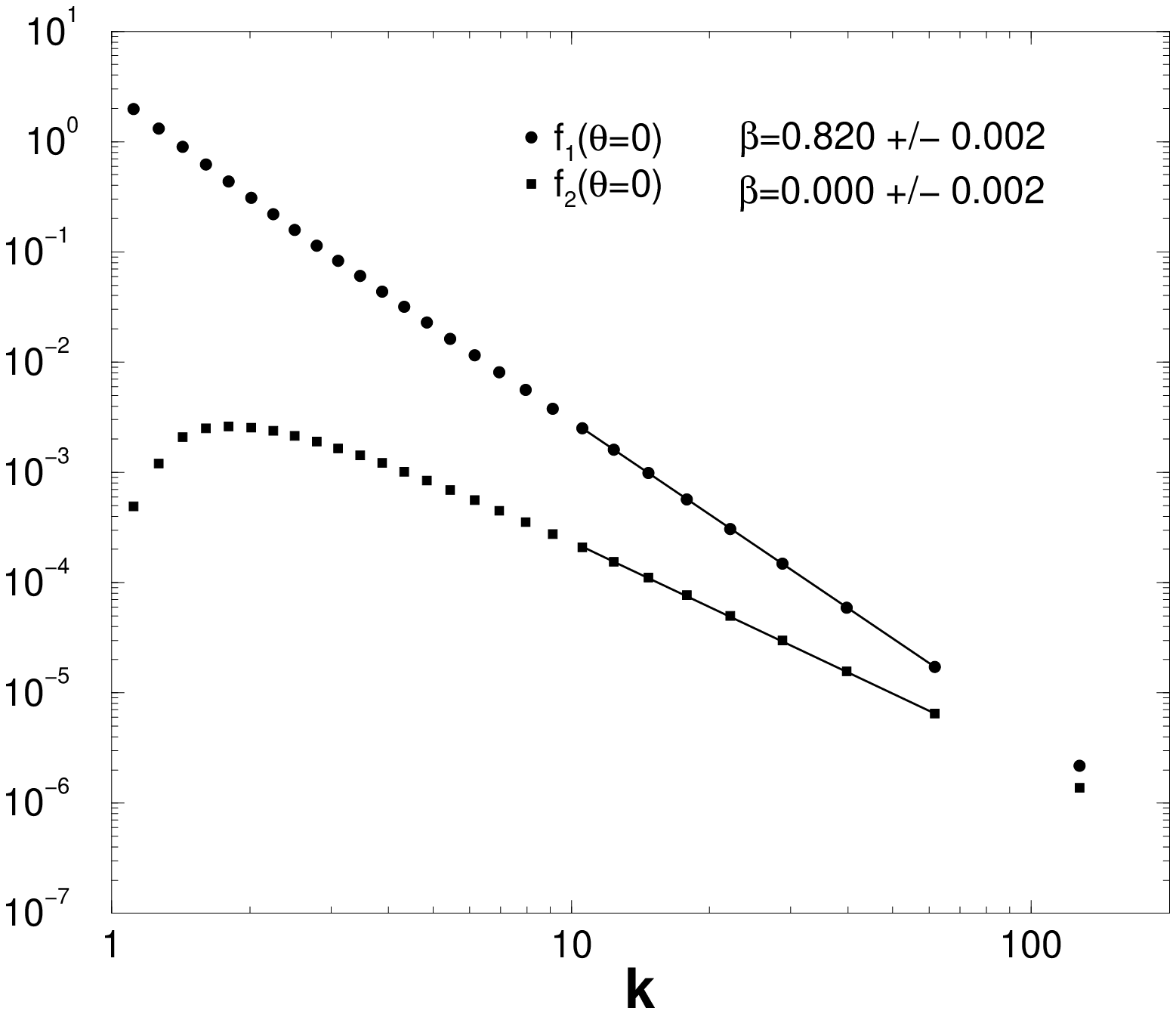}}\end{center}
\vspace{-.5cm}
\caption{Asymptotic behavior of the J=0 wave function 
components $f_i$ for $B$=0.05, $\alpha$=1.096, $\mu$=0.25. The
slope coefficient are $\beta_1=0.82$ and  $\beta_2\approx0$.}\label{wf_as_50}
\end{minipage}
\hspace{0.5cm}
\begin{minipage}[t]{80mm}
\begin{center}\epsfxsize=80mm\mbox{\epsffile{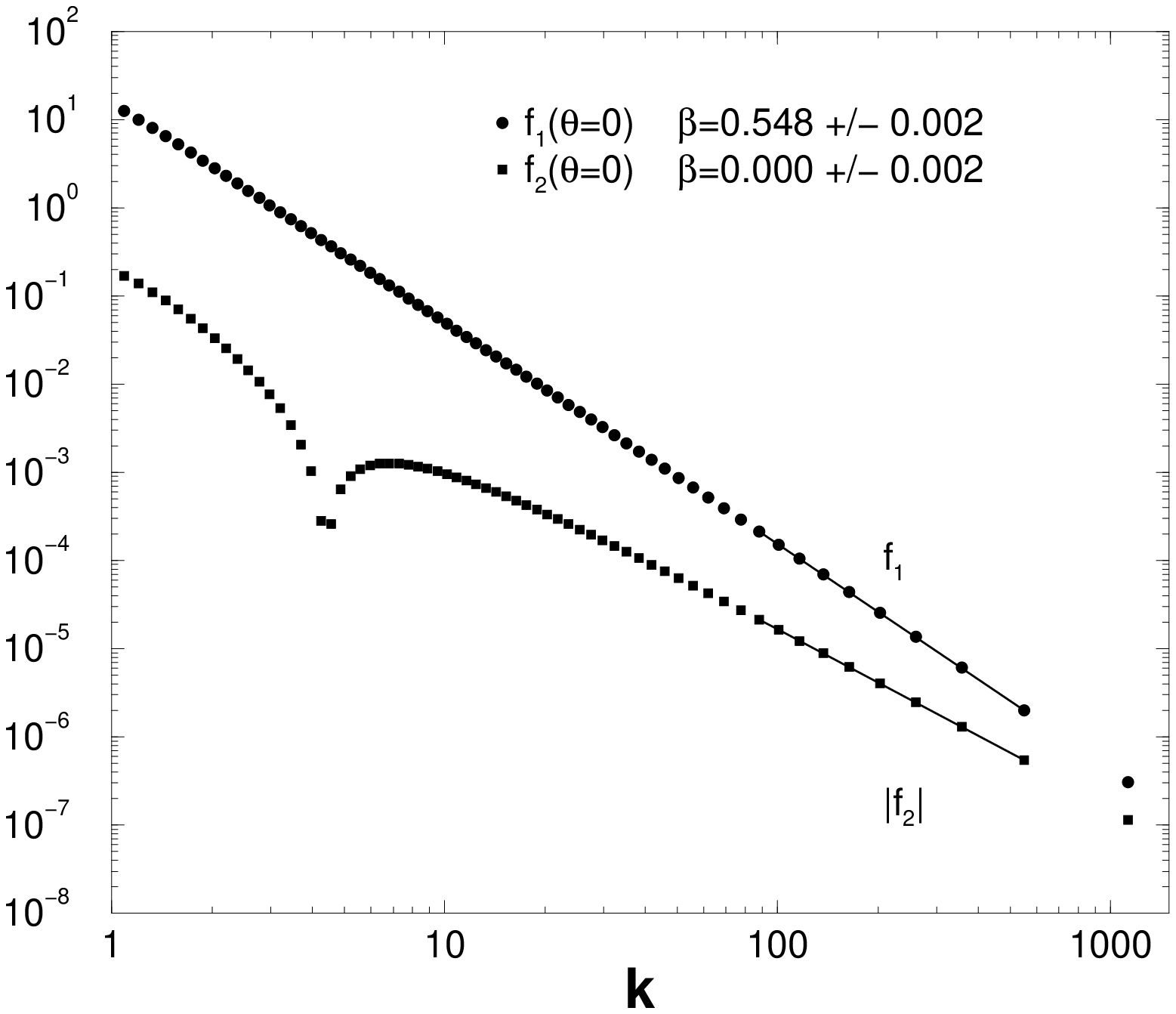}}\end{center}
\vspace{-.5cm}
\caption{Asymptotic behavior of the J=0 wave function 
components $f_i$ for $B$=0.5, $\alpha$=2.48, $\mu$=0.25. The
slope coefficient are $\beta_1=0.55$ and  $\beta_2\approx0$.}\label{wf_as_500}
\end{minipage}
\end{figure}

We have examined the asymptotic behavior of the wave function and found 
that it accurately follows the  power law (\ref{pl}) with a coefficient 
$\beta(\alpha)$ given in Figure \ref{alpha_beta}.  For instance for a 
binding energy B=0.05 ($\alpha=1.096$) a direct measurement in the 
numerical solution  plotted in Figure \ref{wf_as_50} gives 
 $\beta=0.820\pm0.002$ whereas the solution of equation (\ref{Sta_5}) 
for the corresponding $\alpha$ gives $\beta=0.819$.  The same kind of 
agreement was found for B=0.5 ($\alpha=2.480$):  the asymptotic wave 
function -- displayed in Figure \ref{wf_as_500} -- gives 
$\beta=0.548\pm0.002$ and equation (\ref{Sta_5}) provides the value 
$\beta=0.547$.  This agreement shows in particular that the critical 
value of the coupling constant is the same for the one- and the 
two-channel problem.  The influence of the second channel seems to have 
no any effect in the asymptotic behavior of $f_1$.  This channel 
behaves asymptotically  as ${1/k^2}$ i.e. $\beta=0$ for any value of 
the binding energy, as indicated in sect.  \ref{asympt}. One can see 
that the component $f_2$ changes the sign.


It is worth noticing that -- at least in the framework of this model --
one could measure the coupling constant
from the asymptotic behavior of the bound state wave function.
We would like to point out however that the extraction
of coefficient $\beta$ is numerically delicate 
when solving the equation with  finite values of the cutoff $k_{max}$.
A way to overcome this difficulty 
consists in mapping the interval $k\in[0,\infty]$ 
into a compact interval. The mapping $k\rightarrow x={k\over k+m}$ was used.

\begin{figure}[h]
\begin{center}
\mbox{\epsfxsize=9.5cm\epsfysize=9.5cm\epsffile{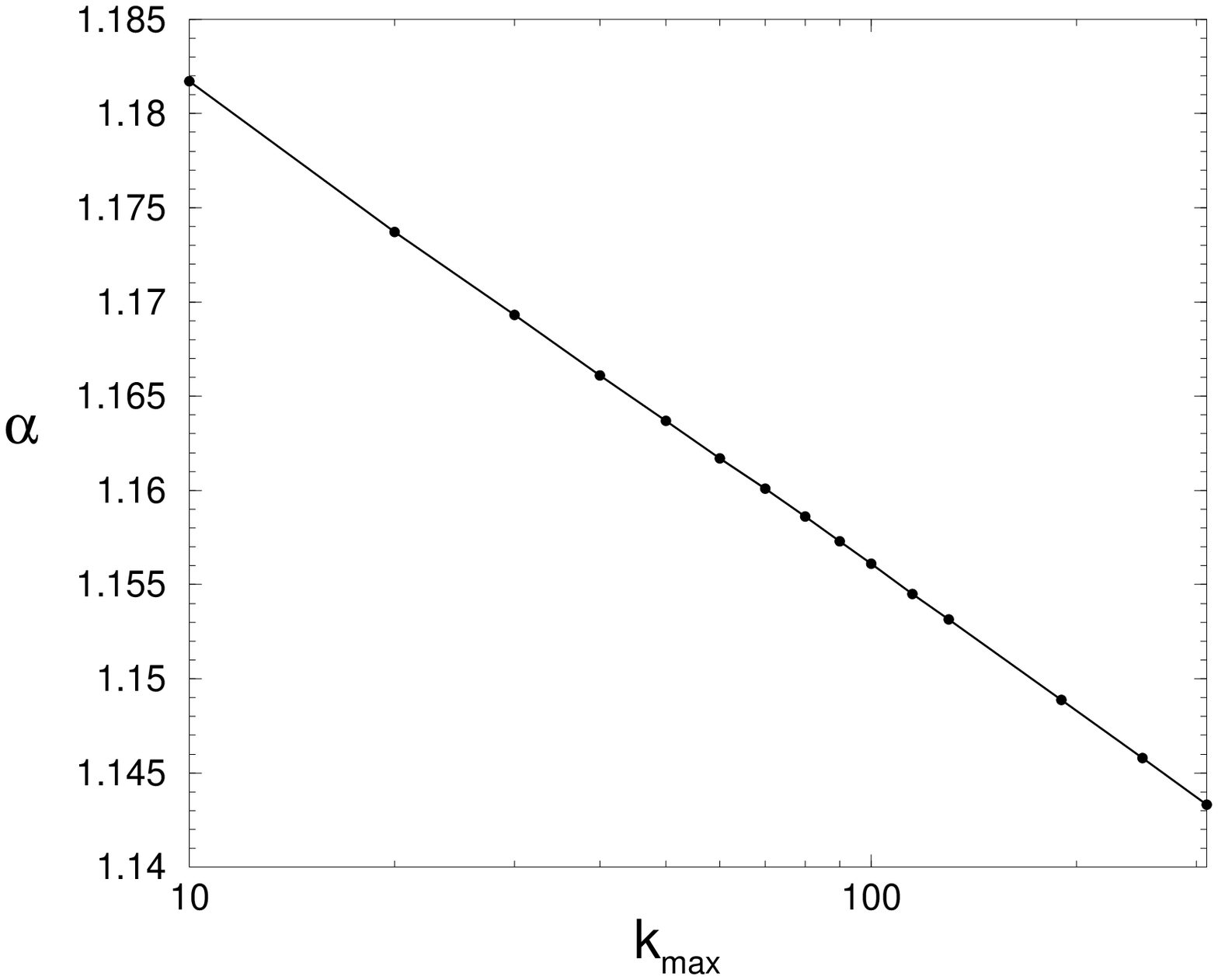}}
\end{center}
\caption{Cutoff dependence of the coupling constant for $J=1,a=1$  
state with $B=0.05$ until $k_{max}$=300. 
It shows a logarithmic divergence.}\label{alpha_kmax_J1a1_50}
\end{figure}

Let us now consider the $J=1$ case. For $J=1,a=0$ or $(1-,2+)$ state, 
contrary to the $J=0$, the value of  $M^2(k_{max})$ -- displayed in Figure
\ref{M2_kmax} -- 
decreases  faster than logarithmically and indicates a collapse.
The asymptotics of  the kernel
$K^{(J=1)}_{22}$  is the same than $-K^{(J=0)}_{22}$, it is negative and
corresponds to attraction. The integral (\ref{H_beta}) for the kernel $H(z,z')$
with the function $A_{22}$ given by (\ref{eq14a_1}) diverges.
Therefore it results in a collapse for any value of the
coupling constant, as pointed out in \cite{glazek1}. 
The same result was found when solving the $J=0$ equations with
the opposite sign of $K^{(J=0)}_{22}$.
The case $J=1,a=1$ state requires a four channel
calculations. The results displayed in Figure \ref{alpha_kmax_J1a1_50}
show a logarithmic divergence.

One can see  from  Figures \ref{alpha_kmax_50} and  
\ref{alpha_kmax_J1a1_50} that the binding energies for the $J=1$ states 
with different values of projection $a$ are different but almost 
degenerate for a wide range of cutoff variation. For instance, with 
$k_{max}=10$ one has $\alpha_{a=0}$=1.17 $\alpha_{a=1}$=1.18 and with 
$k_{max}=90$, $\alpha_{a=0}$=1.14 $\alpha_{a=1}$=1.16.  These 
differences are less than 1\% for a system with not so small binding 
energies (B=0.05) and we expect them to be negligible for weakly bound 
systems like deuteron.  This quasi-degeneracy is much smaller in 
comparison to the results previously found with the scalar particles 
\cite{heidelberg,Taiwan,Miller}.  In this latter case, a bound state 
with the same energy presents a splitting of $\approx20\%$ in $\alpha$, 
what correspond to a energy difference $\Delta B\approx B$. It is worth 
noticing that even for the sizeable cutoff $k_{max}=90$, the value of 
the coupling constant is still $\approx$ 10\% far from the converged 
one obtained by a mapping. 

\begin{figure}[h]
\begin{minipage}[t]{80mm}
\begin{center}
\mbox{\epsfxsize=8.0cm\epsfysize=8.0cm\epsffile{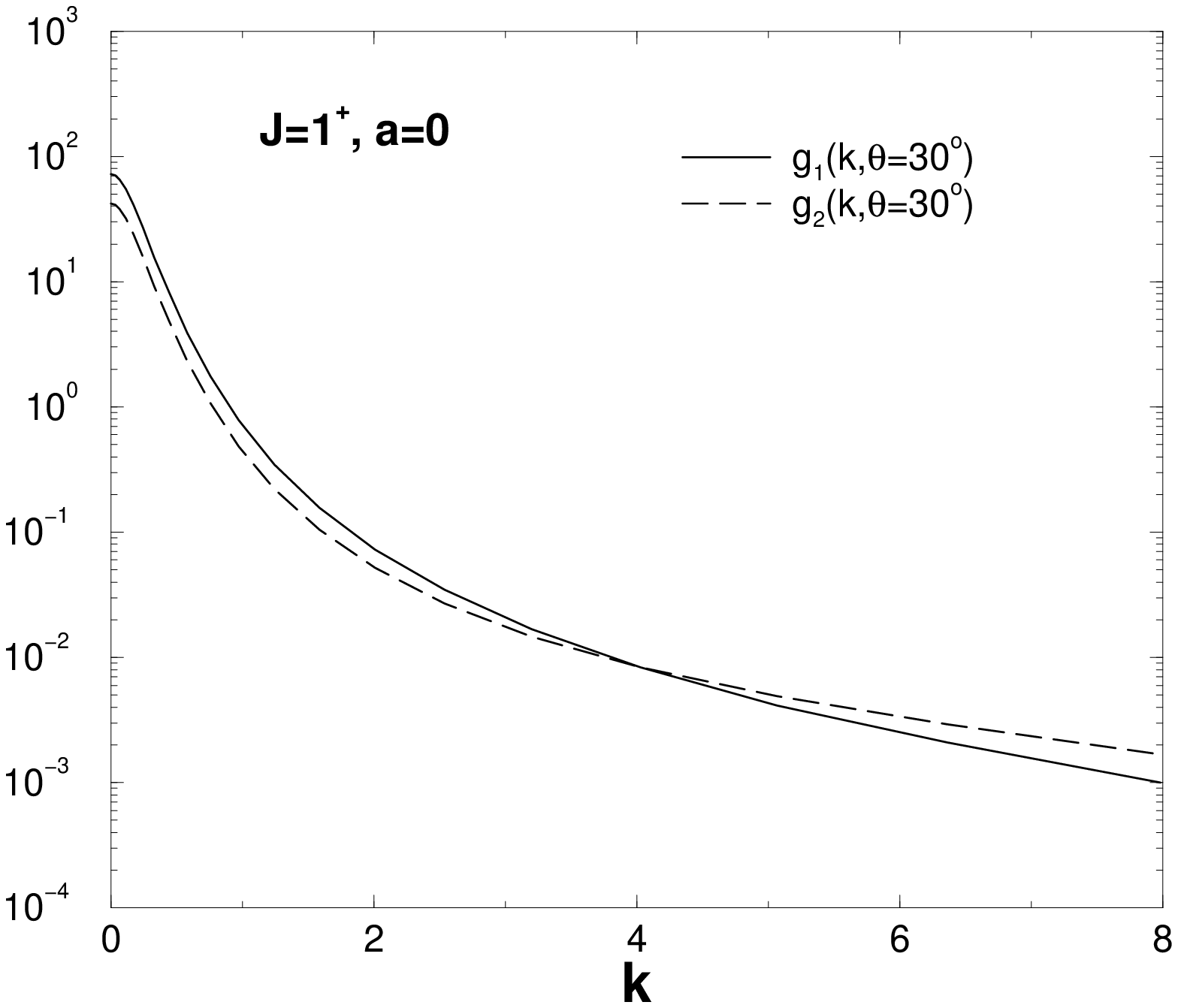}}
\end{center}
\caption{Components $g_1$ and $g_2$ of the $J=1,a=0$ $(1-,2+)$ state, for 
$B=0.05$.} \label{fi_J=1_a=0} 
\end{minipage}\hspace{0.5cm} 
\begin{minipage}[t]{80mm}
\begin{center}
\mbox{\epsfxsize=8.0cm\epsfysize=8.0cm\epsffile{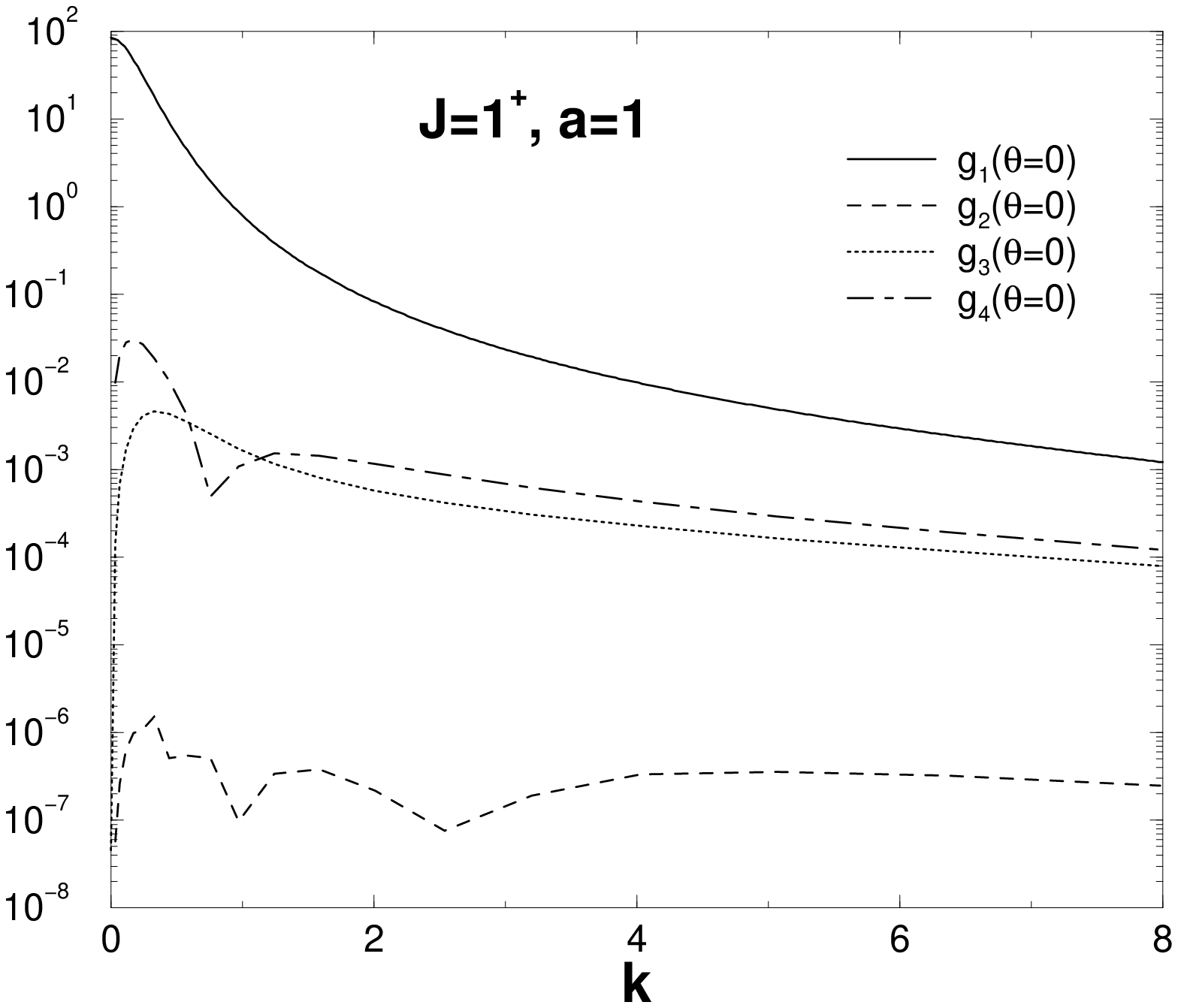}}
\end{center}
\caption{The four components $g_i$ of the $J=1,a=1$ $(1+,2-)$ state, for 
$B=0.05$.}
\label{fi_J=1_a=1}
\end{minipage}
\end{figure}

The 2+4 components of the $J=1^+$ state are displayed in Figures
\ref{fi_J=1_a=0} and \ref{fi_J=1_a=1} as functions of the momentum $k$ . 
Solutions $g_1$ and $g_2$ coming from the $a=0$ sector are comparable. 
They both depend on the angle $\theta$ between the direction of 
the light-front plane and the momentum $\vec{k}$: when $\vec{k}\perp\hat{n}$, 
the first component vanishes
and the second one reaches its maximal value as a function of $\theta$: it 
dominates even over $g_1(k,\theta=0)$ and in the whole range of 
momentum.  In the $J=1,a=0$ state, only one solution is sizeable and 
dominates the three others for all values of $\theta$.  We would like 
to notice that the displayed components are related to the physical 
wave functions considered in \cite{ckj1}
by some specific linear combinations ensuring the correct
transformation properties of the wave function,
as well as an easy link with the non relativistic solutions.
The peculiarities of these
functions and their relations  will be discussed in a forthcoming paper  \cite{fermions}.     

\section{Conclusion}\label{concl}

The Light-Front solutions of the two fermion system interacting via a 
scalar exchange have been obtained. We have found that the $J=0$ -- or 
$(1+,2-)$ -- state is stable (i.e. convergent relative  to the cutoff 
$k_{max}\to \infty$) for coupling constant below some critical value, 
in a way similar to what is known in non relativistic quantum mechanics 
for the $-\alpha'/r^2$ potential.  In this point, our conclusion 
differs from the one settled in \cite{glazek1}, where it was stated 
that the integrals in eqs. (\ref{eq4gl}) diverge logarithmically with 
cutoff.  Above the critical value the system collapses.  This fact 
manifests as an unbounded value of $M^2$ when the cutoff tends to 
infinity. 

We have shown analytically that the asymptotic behavior of the 
wave function has the form ${1\over k^{2+\beta}}$ and that the relation 
between coefficient $\beta$ and the coupling constant can be obtained 
as a solution of an eigenvalue equation suggested by \cite{smirnov}.  
This relation  provides in particular the critical value of the 
coupling constant, which corresponds to $\beta=0$.  These results are 
in agreement with the numerical solutions of the LFD equations.

In the $J=1,a=0$ -- or $(1-,2+)$ -- state the system is found to be 
always unstable, as it was pointed out in  \cite{glazek1}.  The  
instability is related to the dominating $K_{22}$ kernel which is 
attractive.  The origin of the collapse is thus different from $J=0$ 
state, for which the $K_{22}^{(J=0)}$  kernel is repulsive and the 
instability is due to the asymptotic behavior of attractive $K_{11}$
and depends on the value of $\alpha$ relative to $\alpha_c$.

The solutions for the four channel problem $J=1,a=1$ state have been 
obtained.  They show also a logarithmic divergence of $M^2(k_{max})$ 
and are thus unstable without regularization.

These results should be taken into account when carrying out the 
renormalization procedure.  The explicitly covariant version of 
Light-Front Dynamics (CLFD) seems very promising for handling this 
problem \cite{dugne}, like it has proved to be fruitful  in the Yukawa 
model.

\section*{Acknowledgements}
The authors are indebted to St. Glazek and A.V. Smirnov for useful
discussions. One of the authors (V.A.K.) is sincerely grateful for the warm
hospitality of the theory group at the Institut des Sciences Nucl\'{e}aires,
Universit\'e Joseph Fourier, in Grenoble, where this work was performed. 
This work was partially
supported by the grant No. 99-02-17263 of the Russian Fund for Basic
Researches.
The numerical calculations were performed  at CGCV (CEA Grenoble) and  IDRIS
(CNRS). We thank the staff members  of these organizations for their
constant support.

\appendix

\section{Explicit expressions for LFD kernels}\label{scal_1}

The kernels are obtained through the traces (\ref{kappa_ij_J0}). 
To calculate them, we express the scalar products between the 
different four-vectors in terms of  variables $k,k',\theta,\theta',\varphi'$. 
The following results are useful:
\begin{eqnarray*}
&&\omega^2=0\\
&&k_1^2=k_2^2=k'^2_1=k'^2_2=m^2,\nonumber\\
&&\omega\cd k_1=x\; \omega\cd p,\\ 
&&\omega\cd k_2=(1-x)\omega\cd p,\\
&&\omega\cd k'_1=x'\omega\cd p,\\
&&\omega\cd k'_2=(1-x')\omega\cd p,\nonumber\\
&&k_1\cd k_2=2\varepsilon_k^2 - m^2,\\
&&k'_1\cd k'_2=2\varepsilon_{k'}^2 - m^2,\nonumber\\
&&k_1\cd p=2\varepsilon_k^2(1 - x) + M^2 x/2, \\
&&k_2\cd p= 2\varepsilon_k^2 x+M^2 (1 - x)/2 \nonumber\\
&&k'_1\cd p=2 \varepsilon_{k'}^2 (1 - x') + M^2 x'/2, \\
&&k'_2\cd p= 2 \varepsilon_{k'}^2 x'+M^2 (1 - x')/2 ,\nonumber\\
&&k_1\cd k'_1=-kk'\sin\theta\sin\theta'\cos\varphi'+ 2 \varepsilon_{k'}^2 x 
+ 2 \varepsilon_k^2 x' - 
  2 \varepsilon_k^2 x x' - 2 \varepsilon_{k'}^2 x x',
\nonumber\\  
&&k_2\cd k'_2=-kk'\sin\theta\sin\theta'\cos\varphi'
+ 2 \varepsilon_k^2 x + 2 \varepsilon_{k'}^2 x' - 
  2 \varepsilon_k^2 x x' - 2 \varepsilon_{k'}^2 x x',
\nonumber\\ 
&&k_1\cd k'_2=kk'\sin\theta\sin\theta'\cos\varphi'
+ 2 \varepsilon_k^2 (1 - x) (1 - x') + 
  2 \varepsilon_{k'}^2 x x',
\nonumber\\ 
&&k_2\cd k'_1=kk'\sin\theta\sin\theta'\cos\varphi'
+ 2 \varepsilon_{k'}^2 (1 - x) (1 - x') +  2 \varepsilon_k^2 x x',
\end{eqnarray*}
where $2x=\left(1-{k\cos\theta\over\varepsilon_k}\right)$ and similarly for $x'$.

The analytical expressions for the $\kappa_{ij}$ kernels (\ref{kappa_ij_J0})
in the $J=0$ case are:
\begin{eqnarray}\label{eqap1}
\kappa_{11}&=&-\alpha\pi
\left[2 k^2 k'^2+3k^2 m^2+3k'^2 m^2+4 m^4
-2 k k'\varepsilon_k \varepsilon_{k'}\cos\theta \cos\theta'- k k' (\varepsilon_{k}^2+\varepsilon_{k'}^2)
\sin\theta \sin\theta' \cos\varphi'\right]\nonumber\\
\kappa_{12}&=&-\alpha\pi m
(k^2 - k'^2) \left(k'\sin\theta' + k\sin\theta\cos\varphi' \right)\nonumber\\
\kappa_{21}&=&-\alpha\pi m
(k'^2 - k^2) \left(k\sin\theta + k'\sin\theta'\cos\varphi' \right)
\\
\kappa_{22}&=&-\alpha\pi
\left[\left(2 k^2 k'^2+3k^2 m^2+3k'^2 m^2+4 m^4
- 2 k k' \varepsilon_k\varepsilon_{k'}
\cos\theta \cos\theta'\right)\cos\varphi'
-kk'(\varepsilon_{k}^2+\varepsilon_{k'}^2) \sin\theta\sin\theta'\right]\nonumber
\end{eqnarray}
where we have used $\alpha={g^2\over4\pi}$.

The analytical expressions for the $\kappa_{ij}$ kernels
(\ref{eq10b}) in the $J=1^+,a=0$ case are:
\begin{eqnarray}\label{eq11}
\kappa_{11}&=&-\alpha\pi\left\{\left[
-2 k k' \varepsilon_k \varepsilon_{k'}
+(2k^2 k'^2+3m^2k^2+3m^2k'^2 +4m^4)\cos\theta\cos\theta'\right]
+\varepsilon_k \varepsilon_{k'}
(k^2+k'^2+4m^2)\sin\theta\sin\theta'\cos\varphi'\right\}\nonumber\\
\kappa_{12}&=&-\alpha \pi m
\left[-\varepsilon_{k'}(3k^2+k'^2+4m^2)\cos\theta\sin\theta'
+\varepsilon_k ( k^2+3k'^2+4m^2) 
\sin\theta\cos\theta'\cos\varphi'\right]\nonumber\\
\kappa_{21}&=&-\alpha \pi m
\left[-\varepsilon_{k}(k^2+3k'^2+4m^2)\sin\theta\cos\theta'
+\varepsilon_{k'}(3 k^2+k'^2+4m^2) 
\cos\theta\sin\theta'\cos\varphi'\right]\\
\kappa_{22}&=&-\alpha\pi
\left\{\varepsilon_k \varepsilon_{k'}(k^2+k'^2+4m^2)\sin\theta\sin\theta' 
+\left[-2k k'\varepsilon_k \varepsilon_{k'}+
(2k^2 k'^2+3m^2k^2+3m^2k'^2 +4m^4)\cos\theta\cos\theta'\right]
\cos\varphi'\right\}\nonumber
\end{eqnarray}

The analytical expressions for the $\kappa_{ij}$,
determining by eq. (\ref{nz1}) the kernels $K_{ij}$
in the $J=1^+,a=1$ case are:
\begin{eqnarray*}\label{eq11_J1a1} 
 {2\kappa_{11}\over\alpha\pi}  &=&
-m[\varepsilon_{k}\sin^2\theta(\varepsilon_{k}^2
+3\varepsilon_{k'}^2)+\varepsilon_{k'}\sin^2\theta'(3\varepsilon_{k}^2
+\varepsilon_{k'}^2)]
-(\cos^2\theta+\cos^2\theta')[m^2(\varepsilon_{k}^2
+\varepsilon_{k'}^2)+2 \varepsilon_{k}^2\varepsilon_{k'}^2]
+ 4\varepsilon_{k}\varepsilon_{k'}kk'\cos\theta\cos\theta' \cr
&-&(\varepsilon_{k}+\varepsilon_{k'})^2\sin\theta  \sin\theta'   
\cos\varphi' 
\left\{   (\varepsilon_{k}-m)(\varepsilon_{k'}-m) [ \cos\theta\cos\theta'
+\sin\theta\sin\theta'\cos\varphi']-kk' \right\}      \cr
{2\kappa_{12}\over\alpha\pi}&=& 
m\varepsilon_{k}(\varepsilon_{k}^2+3\varepsilon_{k'}^2)\sin^2\theta 
-m\varepsilon_{k'}(3\varepsilon_{k}^2+\varepsilon_{k'}^2)\sin^2\theta'
+ (\cos^2\theta-\cos^2\theta') [m^2\varepsilon_{k}^2+m^2\varepsilon_{k'}^2 
+2\varepsilon_{k}^2\varepsilon_{k'}^2 ]\cr
&-&\sin\theta\sin\theta' \left\{  kk'(\varepsilon_{k}-\varepsilon_{k'})^2
+(\varepsilon_{k}+\varepsilon_{k'})^2(\varepsilon_{k'}-m)(\varepsilon_{k'}-m)\cos\theta\cos\theta 
 \right\} \cos\varphi' \cr
&+&\sin^2\theta(\varepsilon_k-m) \left\{ -(\varepsilon_{k'}+m) 
(\varepsilon_{k}-\varepsilon_{k'})^2+ (\varepsilon_{k'}-m) (\varepsilon_{k}
+\varepsilon_{k'})^2\cos^2\theta'
 \right\}  \cos^2\varphi'   \cr
{2\kappa_{13}\over\sqrt2\alpha\pi}&=&
-\sin\theta'\left[2\varepsilon_{k}\varepsilon_{k'}kk'\cos\theta
+(\varepsilon_{k'}-m)[m(\varepsilon_{k'}^2+\varepsilon_{k}^2)
-2\varepsilon_{k}^2\varepsilon_{k'}]\cos\theta'\right]\cr
&+& \sin\theta \left\{  2\varepsilon_{k}\varepsilon_{k'}kk'\cos\theta'
+(\varepsilon_{k}-m)\cos\theta 
\left[ 
\varepsilon_{k'}(\varepsilon_{k}^2+\varepsilon_{k'}^2-2m\varepsilon_{k})
-(\varepsilon_{k'}-m)(\varepsilon_{k}+\varepsilon_{k'})^2\cos^2\theta'\right] \right\} \cos\varphi' \cr
&-& (\varepsilon_{k}+\varepsilon_{k'})^2 (\varepsilon_{k}-m) 
(\varepsilon_{k'}-m) \sin^2\theta\sin\theta'\cos\theta'  \cos^2\varphi' 
\cr 
{2\kappa_{14}\over\alpha\pi}&=& \sqrt2(k^2-k'^2) \left\{ 
(\varepsilon_{k}\sin^2\theta+m\cos^2\theta)k'\sin\theta'
+mk\sin\theta\cos\varphi'  
-k'(\varepsilon_{k}-m)\sin^2\theta\sin\theta'\cos^2\varphi'  \right\}   \cr
{2\kappa_{21}\over \alpha \pi}&=&  
-m\varepsilon_{k}  \sin^2\theta (3 \varepsilon^2_{k'}+\varepsilon^2_{k})
+m\varepsilon_{k'} \sin^2\theta'(\varepsilon^2_{k'}+3\varepsilon^2_{k}) 
-(\cos^2\theta -\cos^2\theta')\left(m^2(\varepsilon^2_{k'}+\varepsilon^2_{k})
+
2 \varepsilon^2_{k'}\varepsilon^2_{k}\right)\\
&-&\sin\theta \sin\theta' 
\left\{k k' (\varepsilon_{k}-\varepsilon_{k'})^2 
     +(\varepsilon_{k}+\varepsilon_{k'})^2
      (\varepsilon_{k'}-m)(\varepsilon_{k}-m)\cos\theta \cos\theta' \right\} 
\cos\varphi' \\
&+&\sin^2\theta'(\varepsilon_{k'}-m) \left\{ -(\varepsilon_{k'}
-\varepsilon_{k})^2
(\varepsilon_{k}+m)+\cos^2\theta (\varepsilon_{k'}+\varepsilon_{k})^2
(\varepsilon_{k}-m)
\right\} \cos^2\varphi'\\
{2\kappa_{22}\over\alpha\pi}&=& 
m(\varepsilon_{k}+\varepsilon_{k'})^3-(\varepsilon_{k}-m) 
(m(\varepsilon^2_{k}+\varepsilon^2_{k'})
-2\varepsilon_{k}\varepsilon^2_{k'})\cos^2\theta
-(\varepsilon_{k'}-m)(m(\varepsilon^2_{k}+\varepsilon^2_{k'})
-2\varepsilon^2_{k}\varepsilon_{k'})\cos^2\theta' \\
&-&4kk'\varepsilon_{k}\varepsilon_{k'}\cos\theta\cos\theta'
-(\varepsilon_{k}+\varepsilon_{k'})^2  \left\{
(\varepsilon_{k}-m) (\varepsilon_{k'}-m) -k k'
\cos\theta\cos\theta'\right\}\sin\theta\sin\theta' \cos\varphi'\\
&-&  \left\{ \sin^2\theta'
\left(\sin^2\theta \varepsilon_{k}\varepsilon_{k'} 
(\varepsilon^2_{k}+\varepsilon^2_{k'}+2m^2)+
m\varepsilon_{k'} (\cos^2\theta+1)(3\varepsilon^2_{k}
+\varepsilon^2_{k'})\right)
+(\cos^2\theta'+1)\right. \\
&\times& \left.\left(m\varepsilon_{k} \sin^2\theta (\varepsilon^2_{k}
+3\varepsilon^2_{k'})+(\cos^2\theta+1)
(m^2(\varepsilon^2_{k}+\varepsilon^2_{k'})
+2\varepsilon^2_{k}\varepsilon^2_{k'})   \right)
-8k k'\varepsilon_{k}\varepsilon_{k'}\cos\theta\cos\theta' 
\right\} \cos^2\varphi' \\
{2\kappa_{23}\over\sqrt2\alpha\pi}&=&
\left\{ 2 \varepsilon_{k} \varepsilon_{k'} k' k \cos\theta
+(\varepsilon_{k'}-m) \cos\theta' (\varepsilon^2_{k'} m
+\varepsilon^2_{k} (m-2 \varepsilon_{k'})) \right\} \sin\theta'
-\sin\theta\\
&\times&\left\{ 
(\varepsilon_{k}-m)\cos\theta \left(
(\varepsilon_{k}+\varepsilon_{k'})^2(\varepsilon_{k'}-m) 
\cos^2\theta'
-\varepsilon_{k'} (\varepsilon^2_{k}+
\varepsilon^2_{k'}-2m\varepsilon_{k})\right)
-2kk'\varepsilon_{k}\varepsilon_{k'}\cos\theta' \right\}\cos\varphi' \\
&-&\sin\theta' 
\left\{(\varepsilon_{k'}-m)\cos\theta'\left((\varepsilon_{k}
-\varepsilon_{k'})^2
(\varepsilon_{k}+m)-(\varepsilon_{k}+\varepsilon_{k'})^2(\varepsilon_{k}-m)
\cos^2\theta
\right)+4kk'\varepsilon_{k}\varepsilon_{k'}\cos\theta\right\}
\cos^2\varphi' \\
{2\kappa_{24}\over\alpha\pi}  &=&\sqrt{2}(\varepsilon_{k'}^2
-\varepsilon_{k}^2)\left\{mk\sin\theta\cos\varphi'-k'\sin\theta'
\left[(\varepsilon_{k}\sin^2\theta+m\cos^2\theta)-[(\varepsilon_{k}
+m)-(\varepsilon_{k}-m)\cos^2\theta]\cos^2\varphi'\right]
\right\}  \cr
&&\cr
{2\kappa_{31}\over\sqrt2\alpha\pi} &=& -\sin\theta\left[ (\varepsilon_{k}-m)
(m\varepsilon_{k}^2+m\varepsilon_{k'}^2-2\varepsilon_{k}\varepsilon_{k'}^2)
+2 \varepsilon_{k}\varepsilon_{k'}kk' \cos\theta' \right]\cr
&+& \sin\theta'\left\{   (\varepsilon_{k}-m)\cos\theta'
\left[ (m\varepsilon_{k}^2+m\varepsilon_{k'}^2
-2\varepsilon_{k}^2\varepsilon_{k'})\cos^2\theta 
+\varepsilon_{k}(\varepsilon_{k}^2+\varepsilon_{k'}^2
-2m\varepsilon_{k'})\sin^2\theta\right] +2\varepsilon_{k}
\varepsilon_{k'}kk'\cos\theta' \right\}\cos\varphi'  \cr
&-&  (\varepsilon_{k}+\varepsilon_{k'})^2 (\varepsilon_{k}-m) 
(\varepsilon_{k'}-m) \sin\theta\cos\theta\sin^2\theta'\cos^2\varphi'\cr
&&\cr
{2\kappa_{32}\over\sqrt2\alpha\pi}&=&  \sin\theta\left[2\varepsilon_{k}
\varepsilon_{k'} kk'\cos\theta' +(e-m)[m(\varepsilon_{k}^2
+\varepsilon_{k'}^2)-2\varepsilon_{k}\varepsilon_{k'}^2]
\cos\theta \right]  \cr
&+& \sin\theta'\cos\varphi'\left\{(\varepsilon_{k'}-m)\cos\theta'
\left[ 
\cos^2\theta(m\varepsilon_{k}^2+m\varepsilon_{k'}^2
-2\varepsilon_{k}^2\varepsilon_{k'}^2) + \varepsilon_k\sin^2\theta
[\varepsilon_{k}^2+\varepsilon_{k'}^2-2m\varepsilon_{k'}\right]
+2\varepsilon_{k}\varepsilon_{k'}kk'\cos\theta  \right\}  \cr
&+&  \sin\theta\cos^2\varphi'\left\{ 
(\varepsilon_{k}-m)\cos\theta \left[ \varepsilon_{k'}
(2\varepsilon_{k}m-\varepsilon_{k}^2-\varepsilon_{k'}^2)\sin^2\theta'   
   + (2\varepsilon_{k}\varepsilon_{k'}^2-m\varepsilon_{k}^2
   -m\varepsilon_{k'}^2) (1+\cos^2\theta') \right]
  -4\varepsilon_{k}\varepsilon_{k'} kk' \cos\theta' \right\} \cr
&&\cr
 {\kappa_{33}\over\alpha\pi}  &=& \left\{
\varepsilon_{k} (\varepsilon_{k'}-m)\left(\varepsilon^2_{k} 
+\varepsilon_{k'} (\varepsilon_{k'}-2m)\right)
\cos^2\theta' 
-\varepsilon_{k'}m(3\varepsilon^2_{k}+\varepsilon^2_{k'})\cos^2\theta
-\varepsilon_{k}\varepsilon_{k'} 
(\varepsilon^2_{k}+\varepsilon^2_{k'}+2m^2)
\sin^2\theta \right.\\
&-&\left.\left((\varepsilon_{k}+\varepsilon_{k'})^2
(\varepsilon_{k}-m)(\varepsilon_{k'}-m)\cos\theta\cos\theta'
- 2 k k'\varepsilon_{k}\varepsilon_{k'}\right)
(\cos\theta\cos\theta'+\sin\theta \sin\theta' \cos\varphi')\right\}
\cos\varphi' \\
&&\cr
{\kappa_{34}\over\alpha\pi}&=& (\varepsilon_{k}^2-\varepsilon_{k'}^2) 
(\varepsilon_{k}-m)\cos\theta\sin\theta k'\sin\theta' \sin^2\varphi' \cr
{2\kappa_{41}\over\alpha\pi}&=&\sqrt2 (\varepsilon_{k'}^2
-\varepsilon_{k}^2)  
\left\{   mk'\sin\theta'\cos\varphi'  
+  k\sin\theta \left[(\varepsilon_{k'}\sin^2\theta'+m\cos^2\theta')
-(\varepsilon_{k'}-m)\sin^2\theta'\sin^2\varphi'\right] \right\} \cr
{2\kappa_{42}\over\alpha\pi}&=&\sqrt2 (\varepsilon_{k}^2
-\varepsilon_{k'}^2)
\left\{   mk'\sin\theta'\cos\varphi'  
+  k\sin\theta  \left[  [(m+\varepsilon_{k'})
+(m-\varepsilon_{k'})\cos^2\theta']\cos^2\varphi'     
- (\varepsilon_{k'}\sin^2\theta'+m\cos^2\theta') \right]\right\}  \cr
{\kappa_{43}\over\alpha\pi}&=&-k(\varepsilon_{k}^2-\varepsilon_{k'}^2)
(\varepsilon_{k'}-m)\cos\theta'\sin\theta\sin\theta'\sin^2\varphi'\cr
&&\cr
{\kappa_{44}\over\alpha\pi}&=&\left\{ 
  (\varepsilon_{k}^2+\varepsilon_{k'}^2) (kk' \sin\theta\sin\theta'
  \cos\varphi' -m^2)
+ 2\varepsilon_{k}\varepsilon_{k'} ( kk' \cos\theta\cos\theta'  
- \varepsilon_{k}\varepsilon_{k'}) \right\}  \cos\varphi'  
\end{eqnarray*}

\section{Coefficients $\eta_{ij}$ determining the
orthonormalized spin structures}\label{app3}

The construction method   of the four orthonormalized spin structures 
$S^{(1)}_{i\mu}$ (\ref{nzf2}) determining the wave function 
$\phi^{(1)}_{\mu}$ (\ref{nz2_1}), is explained in \cite{These_MMB}. These 
structures are expressed in terms of six structures $S_{j\mu}$, in the 
form (\ref{nzf2}).  The non zero coefficients $\eta_{ij}$ 
($i=1,\ldots,4;\; j=1,\ldots,6$) are given below ($z=\cos\theta$):  
\begin{eqnarray}\label{hij}
&&\eta_{11}=\frac{\sqrt{3}m^2}{4\varepsilon_k(\varepsilon_k+m)},\quad
\eta_{12}=\frac{\sqrt{3}m}{4\varepsilon_k},\quad
\eta_{13}=-\frac{\sqrt{3}(\varepsilon_k-m)(4\varepsilon_k^2+M^2)z}
{16 \varepsilon_k^2 k},
\nonumber\\
&&\eta_{16}=\frac{\sqrt{3}(4\varepsilon_k^2+M^2)}{8\varepsilon_k m},
\nonumber\\
&&\eta_{21}=\frac{\sqrt{3}m^2[\varepsilon_k(1-z^2)+m(1+z^2)]}
{4\varepsilon_k k^2(1-z^2)},\quad
\eta_{22}=-\frac{\sqrt{3}m}{4\varepsilon_k},
\nonumber\\
&&\eta_{23}=-\frac{\sqrt{3}(4\varepsilon_k^2+M^2)
[\varepsilon_k(1-z^2)+m(1+z^2)]z}
{16\varepsilon_k^2 k(1-z^2)},\quad
\eta_{24}=\frac{\sqrt{3}mz}{k(1-z^2)},
\nonumber\\
&&\eta_{26}=-\frac{\sqrt{3}(4\varepsilon_k^2+M^2)(1+z^2)}
{8\varepsilon_k m (1-z^2)}
\nonumber\\
&&\eta_{31}=\frac{\sqrt{3}m^2 z}
{2\varepsilon_k(\varepsilon_k+m)\sqrt{2(1-z^2)}}, \quad
\eta_{33}=-\frac{\sqrt{3}(\varepsilon_k-m)(4\varepsilon_k^2+M^2)z^2}
{8\varepsilon_k^2 k \sqrt{2(1-z^2)}},
\nonumber\\
&&\eta_{34}=-\frac{\sqrt{3}m}{k\sqrt{2(1-z^2)}}, \quad
\eta_{36}=\frac{\sqrt{3}(4\varepsilon_k^2+M^2)z}
{4\varepsilon_k m \sqrt{2(1-z^2)}} \nonumber\\
&&\eta_{45}=\frac{\sqrt{3}m^2}{2\varepsilon_k k \sqrt{2(1-z^2)}}
\end{eqnarray}


\end{document}